\documentclass{elsart}
\usepackage{mathrsfs}
\usepackage{amsmath}
\usepackage{ulem}
\usepackage{soul}
\setstcolor{red}

\usepackage{color}
\usepackage{graphics}
\usepackage{graphicx}
\usepackage{epsfig}

\usepackage{amssymb}

 \usepackage{lineno}

\usepackage{hyperref}
\newtheorem{corollary}{Corollary}

\begin{document}
\bibliographystyle{plainnat}


\begin{frontmatter}
 {\center\title{A Lifting Relation from Macroscopic Variables to Mesoscopic Variables in Lattice Boltzmann Method: Derivation, Numerical Assessments and Coupling Computations Validation }
 \author[xjtu,upmc]{Hui Xu\ead{xuhuixj@gmail.com or xu@lmm.jussieu.fr}},
  \author[xjtu]{Huibao Luan},
  \author[xjtu]{Yaling He},
  \author[xjtu]{Wenquan Tao \corauthref{cor1}\ead{wqtao@mail.xjtu.edu.cn}}
  \corauth[cor1]{Corresponding author. Tel:+86-29-82669106. Fax:+86-29-82669106.}
  \address[xjtu]{Key Laboratory of Thermo-Fluid Science $\&$ Engineering, Xi'an Jiaotong University, Xi'an, Shaanxi, 710049, China.}
  \address[upmc]{Institut Jean le Rond d'Alembert, UMR CNRS 7190, Universit\'e Pierre et Marie Curie - Paris 6, 4 Place Jussieu case 162 Tour 55-65, 75252 Paris
Cedex 05, France}
  }


\end{frontmatter}
\newpage
\section*{Abstract}
In this paper, analytic relations between the macroscopic variables
and the mesoscopic variables are derived for lattice Boltzmann
methods (LBM). The analytic relations are achieved by two different
methods for the exchange from velocity fields of finite-type methods
to the single particle distribution functions of LBM. The numerical
errors of reconstructing the single particle distribution functions
and the non-equilibrium distribution function by macroscopic fields
are investigated. Results show that their accuracy is better than
the existing ones. The proposed reconstruction operator has been
used to implement the coupling computations of LBM and
macro-numerical methods of FVM. The lid-driven cavity flow is chosen
to carry out the coupling computations based on the numerical
strategies of domain decomposition methods (DDM). The numerical
results show that the proposed lifting relations are accurate and
robust.

Key words: LBM, Navier-Stokes equations, non-equilibrium
distribution functions, multi-scale perturbation expansion, coupling
computation, FVM

\newpage
\section{Introduction}
\label{Intro}In the past decades, LBM has been widely used to
simulate fluid flow problems \cite{BenziSucci,Succi}, including
complex turbulent fluid flows \cite{Orszag,Chen1} and multiscale
modeling\cite{kang,Fyta}. This method is based on the Boltzmann
kinetic equation which is used to describe a number of interacting
populations of particles. As described in \cite{SucciT}, ``The LBE
could potentially play a twofold function-as a telescope for the
atomistic scale and a microscope for the macroscopic scale". In
\cite{DUpuis} dense fluids flow past and through a carbon nano tube
(CNT) was studied by a hybrid model coupling LBM and MDS. The
authors pointed out that replacing the finite volume solver by a LBM
aims to take advantage of the mesoscopic modeling inherent in LB
simulations. Thus LBM is a mesoscopic method in nature is a
widely-accepted understanding in the literature. The macroscopic
parameters such as fluid density, velocity and pressure can be
obtained via some averages of the mesoscopic variable which conform
the basic conservation laws of mass and momentum \cite{Succi}. In
practical applications of LBM to simulate a macroscopic problem, a
crucial problem is confronted, that is, a reasonable initial
meso-field must be specified to start the evolution process. The
first initializing method was proposed in \cite{Skordos} in 1993.
Recently, several methods have been proposed to improve the accuracy
of numerical results and reduce the initial layers (oscillation
layers) \cite{Caiazzo,Mei}. Such oscillations have a numerical
origin and are due to the artificial compressibility of LBM. Here,
`` initial layer " refers to such a computational stage within which
the macroscopic parameters are oscillating. When the initial data is
not well-prepared, there is an initial layer during which the
solution adapts itself to match the profile dictated by the
environment. For the LBM, the existence of the initial layers is a
common phenomenon \cite{Caiazzo}. In this paper, we will derive the
lifting relations between the macroscopic variables and the
mesoscopic variables in LBM by two ways. According to the authors'
knowledge, the proposed lifting relations in this paper are
different from those in the existing literature
\cite{Skordos,Caiazzo,Mei,Mohamad,junk1,junk2,junk3}. The proposed
relations will offer us some new views about the reconstruction of
nonequilibrium distribution functions in LBM.

 Challenging multiscale phenomena or processes are widely existed in material science,
chemical engineering process, energy and power engineering, and
other engineering fields. Generally speaking, for a multiscale
problem, we often must use different methods to
numerically model the processes at different geometric
sub-regions and exchange solution information at interface
\cite{Connel,Abraham,Liu,Chopard}. Such coupling computations are
widely adopted in the present-day multiscale simulation. As
indicated above LBM is a kind of mesoscopic methods, which is a
candidate to implement the meso-macro or micro-meso coupling
computations in engineering applications \cite{SucciT}. So, the
proposed method not only can be used to obtain a better initial
field for LBM, but also can be adopted in the multi-scale
computation. For example in \cite{SucciT} the possibility of
coupling LBM with molecular dynamics simulation (MDS) was
investigated and found that with proper time and geometric scales
the two numerical methods can be coupled. And in \cite{DUpuis} such
coupling simulation was conducted. In the existing literatures the
coupling of finite difference method (FDM, which is a macrosopic
method) with LBM was adopted in \cite{Chopard,Leemput,Leemput1}, but the
proposed coupling method is similar to a multigrid method and a
simple regularization formula is used in their computations. The
regularization formula in \cite{Chopard} only considers the
first-order approximation of the single particle distribution
function and the coupling formula in \cite{Leemput} is only used to
deal with the one-dimensional reaction-diffusion system. In
\cite{DUpuis} the coupling between LBM and MDS was implemented by
exchange of velocity and velocity gradient at the interface region.
In this paper, the proposed meso-macro (or micro-meso) coupling
is expected to be used for domain decomposition methods, in which
LBM and macro-type numerical method (or micro-type numerical method
and LBM) are adopted in different sub-domain and information is
exchanged at the interface. We believe that our proposed relation is
more useful method for engineering multiscale computations. In
addition, the proposed coupling method can also be used to carry out
the multigrid computations and equation-free multiscale (EFM)
computations \cite{Kevrekinids}.  It is well-known that LBM is very powerfull for  the parallel computing on a low cost \cite{Mazzeo,Amati}. So, the proposed relation can be used in the parallel simulations for multiscale simulations of complex fluid flows based on the refinement strategies.

To the authors' understanding the glossary ``lifting relation''
means that macroscopic variables in a lower degree-of-freedom (DoF)
system are upscaled to meso/microscopic variables in a higher DoF
system. Generally, it is difficult to establish the one-to-one map
from a lower DoF system to a higher DoF system, although the lower
DoF system can be seemed to be an approximate or approaching form of
a higher DoF system in some referred scales. This situation happens
when numerical results of different scales are coupled at the same
location. For example when MDS and continuum method are coupled,
reference \cite{NIE} indicated that it is straightforward to obtain
the continuum quantities (such as velocity, pressure) from the
particle description by averaging over the local region and over
time, but the reverse problem, generating meso/microscopic particle
configuration from known macroscopic quantities is non-trivial and
must necessarily be non-unique. The glossary ``lifting relation'' in
the title of this paper is proposed based on the concept of the DoF
of the governing equations.

In this paper, we will give two methods to establish the relations
between variables of the Navier-Stokes equations and variables of
LBM. Numerical tests demonstrate that the proposed methods of
computing non-equilibrium distribution functions are effective and
accurate.

The rest of the paper is organized as follow. In section \ref{LBH},
the details of multi-scale derivation of non-equilibrium
distribution functions is given. In section \ref{IS}, the
non-equilibrium distribution functions are obtained by Boltzmann-BGK
equations. In section \ref{NT}, the
 performances of the proposed relations to reconstruct
non-equilibrium distribution functions are demonstrated by numerical
tests. Finally, some conclusions are given.

\section{Lattice Boltzmann hydrodynamics and multiscale approach} \label{LBH}
In this section, we will review LBM and the corresponding
macroscopic equation. Based on this review, we will derive a
relation for lifting macroscopic variables to microscopic variables
by multiscale approach.

\subsection{Lattice Boltzmann hydrodynamics }
 We now introduce the lattice Boltzmann-BGK model as a solver for the weakly-compressible Navier-Stokes equations. LBM is built up from the
lattice gas cellular automata models \cite{Succi}. The numerical
scheme of LBM is established based on a finite discrete-velocity
model of the Boltzmann-BGK equation and can be expressed as follows
\begin{equation}\label{lbm}
f_i({\rm{x}}+\delta t{\rm{c}}_i,t+\delta t)-f({\rm x},t)=\Omega_i,
\end{equation}
where $f_i$ represents the single-particle distribution function
along the direction ${\rm c}_i$ ( $i=0,\ldots,n$), ${\rm c}_i$ is
the element of the discrete velocity set $\mathcal {V}=\{{\rm c}_0,
\ldots, {\rm c}_n\}$. $\Omega_i$ denotes the collision operator
which is non-dimensional. The macroscopic variables,
the density $\rho$ and the velocity ${\rm u}$, are defined locally
by the distribution functions as follows
\begin{equation}\label{macrorho}
\rho({\rm x},t)=\sum_{i=0}^{n}f_i({\rm x},t)=\sum_{i=0}^{n}f_i^{\rm
eq}({\rm x},t),
\end{equation}
\begin{equation}\label{velocity}
{\rm u}({\rm x},t)=\frac{1}{\rho}\sum_{{\rm c}_i\in \mathcal {V}}
{\rm c}_if_i({\rm x},t)=\frac{1}{\rho}\sum_{{\rm c}_i\in \mathcal
{V}} {\rm c}_if_i^{\rm (eq)}({\rm x},t).
\end{equation}
 For the standard LBM, the collision operator is defined by the
 so-called BGK collision
 \begin{equation}
 \Omega_i^{\rm BGK}=-\frac{1}{\tau_{\rm lbm}}[f_i({\rm x},t)-f_i^{\rm (eq)}({\rm x},t)].
 \end{equation}
 For the convenience of comparison, from here, we use the similar notations in \cite{MJun2}. The local equilibrium distribution $f_i^{\rm(eq)}$ is defined by
 \begin{equation}\label{equilibrium}
 f_i^{\rm(eq)}({\rm x},t)=f_i^{L(eq)}({\rm x},t)+f_i^{\rm Q(eq)}({\rm x},t),
 \end{equation}
 where $f_i^{\rm L(eq)}({\rm x},t)$ and
 $f_i^{\rm Q(eq)}({\rm x},t)$ denote the linear part and the
 quadratic part
 of the equilibrium distribution, respectively. The linear part is
 given by
 \begin{equation}\label{leq}
 f_i^{\rm L(eq)}({\rm x},t)=\omega_i\rho(1+\frac{1}{c_s^2}{\rm c}_i\cdot
 {\rm u}({\rm x},t)),
 \end{equation}
 and the quadratic part is expressed by
 \begin{equation}\label{qeq}
 f_i^{\rm Q(eq)}({\rm x},t)=\omega_i\frac{1}{2c_s^4}\rho({\rm u}({\rm x},t){\rm u}({\rm x},t)):\Sigma_i,
 \end{equation}
 where $c_s$ is the lattice sound speed of the model, $\omega_i$ denotes the weight and $\Sigma_i$ is a
 second-order tensor defined by
 \begin{equation}
 \Sigma_{i\alpha\beta}=c_{i\alpha}c_{i\beta}-c_s^2\delta_{\alpha\beta}.
 \end{equation}
The tensor product definition between two first order tensors
$\mathbf{a}$ and $\mathbf{b}$ is given as follows
 \begin{equation}
 (\mathbf{a}\mathbf{b})_{\alpha\beta}=\mathbf{a}_\alpha\mathbf{b}_\beta,
 \end{equation}
 and the corresponding second-order tensor $:$-product between $\mathbf{A}$ and $\mathbf{B}$
 is given by
 \begin{equation}
 \mathbf{A}:\mathbf{B}=\sum_{\alpha,\beta=1}^d \mathbf{A}_{\alpha\beta}
 \mathbf{B}_{\alpha\beta},
 \end{equation}
 where $d$ denotes the spatial dimension.

 In this paper, we mainly focus on the standard LBM. By the Chapman-Enskog
 expansion, under the small $Ma$ number restriction ($Ma\leq0.2)$, we can recover the Navier-Stokes equations as
 follows
 \begin{equation}\label{rho}
 \partial_t{\rho}+\partial_\alpha(\rho u_\alpha)+\mathbf{O}(\delta
 t^2)=0,
 \end{equation}
 \begin{equation}\label{moment}
 \partial_t(\rho u_\alpha)+\partial_\beta(\rho u_\alpha
 u_\beta)=-\partial_\alpha p+\nu\partial_\beta(\rho(\partial_\alpha u_\beta+\partial_\beta
 u_\alpha))+\mathbf{O}(\delta
 t^2)+\mathbf{O}(\delta t u^3),
 \end{equation}
 where $p$ is defined by
 \begin{equation*}
 p=c_s^2\rho.
 \end{equation*}
 It is clear that the recovered Navier-Stokes equations are weakly
 compressible \cite{Succi,ChenDoolen,RicotM}.
 So, the density is coupled with the pressure field in LBM. In Eq. (\ref{moment}), the second term of R.H.S can be rewritten as
 \begin{equation}\label{viscosity}
 \nu\partial_\beta(\rho(\partial_\alpha u_\beta+\partial_\beta
 u_\alpha))=\nu\rho(\partial_\beta\partial_\beta
 u_\alpha)+\nu(\partial_\beta\rho)(\partial_\alpha u_\beta+\partial_\beta
 u_\alpha)+\nu\rho\partial_\alpha\partial_\beta u_\beta.
 \end{equation}
And the corresponding third-order term
$\mathbf{O}(\delta t u^3)$ is given by
\begin{equation}
\mathbf{O}(\delta tu^3)=-\sigma\partial_\beta\partial_\gamma(\rho
u_\alpha u_\beta u_\gamma).
\end{equation}
The fluid viscosity $\nu$ is defined by
\begin{equation}
\nu=c_s^2(\tau_{\rm lbm}-\frac{1}{2})\delta t,
\end{equation}
and $\sigma$ is given by
\begin{equation}
\sigma=\frac{\nu}{c_s^2}.
\end{equation}
In Eq. (\ref{viscosity}), the third term of R.H.S will vanish for a
divergence-free field. But the second term will not vanish, if the
density $\rho$ is nonhomogeneous in the spatial domain. The
Navier-Stokes equations are recovered by LBM under the low Mach
condition. Physically, LBM is a weakly compressible model for
solving Navier-Stokes equations.

At this point, we describe two situations where the lifting relation
is useful. The first situation is using the lifting
relation to get a good initial field of the density distribution
function from specified velocity and pressure fields. As indicated
above the recovered Navier-Stokes equations are weakly compressible,
hence pressure field is coupled with the density field by the equation
of state ($p=c_s^2\rho$). In engineering computations, the
weakly-compressible flow is often used as an approximation of the
incompressible flow. For the lifting function, the consideration
should be made from the weakly compressible side. The
non-homogeneous character of the initial density is very significant
for an initial routine of LBM in the proposed lift relation. This
significance can be observed from the follow-up derivations. For the
initial processes, if the initial pressure field is given, the
lifting relation can be used to obtain the initial distribution
functions consistent with the recovered Navier-Stokes equations. In
another development when we couple LBM with other macroscopic solver
of Navier-Stokes equations, we need to pass the macroscopic
variables (pressure and velocity fields) to an approximate single
particle distribution functions or the non-equilibrium distribution
functions. At this time, a macroscopic equation relating to the
given velocity and pressure to the particle distribution function of
LBM become very useful. The major goal of the present paper is to
derive such a lift relation, or a reconstruction operator as
depicted in \cite{Skordos}.

For the convenience of deriving such an equation, some changes are
made for the form of Eq. (\ref{moment}). We first rewrite Eq.
(\ref{moment}) as
\begin{equation}\label{c}
\begin{array}{c}
\partial_t(\rho u_\alpha)+\partial_\beta(\rho u_\alpha
 u_\beta)=\rho(\partial_t u_\alpha+u_\beta\partial_\beta u_\alpha
 )+u_\alpha (\partial_t \rho+\partial_\beta(\rho u_\beta)).
 \end{array}
\end{equation}
If the initial velocity field is divergence-free, we have
\begin{equation}
\partial_t(\rho u_\alpha)+\partial_\beta(\rho u_\alpha
 u_\beta)=\rho(\partial_t u_\alpha+\partial_\beta(u_\alpha
 u_\beta))+u_\alpha (\partial_t \rho+\partial_\beta(\rho u_\beta)).
\end{equation}
The neglecting of the term $\rho u_\alpha\partial_\beta u_\beta$ is
a widely accepted approximation.  According to Eq. (\ref{rho}), we
have
\begin{equation}\label{estimate}
\partial_t(\rho u_\alpha)+\partial_\beta(\rho u_\alpha
 u_\beta)=\rho(\partial_t u_\alpha+\partial_\beta(u_\alpha
 u_\beta)).
\end{equation}
Now, combining Eq. (\ref{moment}), Eq. (\ref{viscosity}) and Eq.
(\ref{c}), we gain
\begin{equation}\label{finalNSE}
\begin{array}{c}
\partial_t u_\alpha+u_\beta\partial_\beta u_\alpha
 =-\frac{\partial_\alpha p}{\rho}+\nu(\partial_\beta\partial_\beta
 u_\alpha+\partial_\alpha\partial_\beta u_\beta)+\nu\frac{\partial_\beta \rho}{\rho}(\partial_\alpha u_\beta+\partial_\beta
 u_\alpha).
 \end{array}
\end{equation}

\subsection{Derivation of Non-equilibrium Distribution Function by Multi-scale Approach}

The coupled macro-micro/mesoscale simulation is a rapidly developing
area of research that deals with processes covering several order of
geometries. For such numerical approach, one needs to construct an
initial condition $u(x,0)$ for the meso/microscopic simulator, which
is corresponding to the initial macroscopic variable $U(x,0)$. Here,
$u(x,0)$ represents the meso/microscopic state variables and
$U(x,0)$ stands for macroscopic state variables. As indicated above
this procedure is called $lifting$ \cite{Kevrekinids} or
$reconstruction$ \cite{Weinan} step. The lifting (reconstruction)
operator $\mu$ is defined by
\begin{equation}
u(x,0)=\mu(U(x,0)).
\end{equation}
The lifting procedure leads to a one-to-many mapping. After the
initialization of the meso/microscopic variables by
the reconstruction operator $\mu$, they will be evolved by the
meso/microscopic simulator. In this paper, LBM is adopted as the
mesoscopic simulator. As indicated in \cite{Liu,Leemput} the
macroscopic state variables are easy to be achieved. To transfer the
micro/meso-scale parameters into macro parameters we need some
$restriction$ \cite{Kevrekinids} or $compression$ \cite{Weinan}
operators. Conceptually, this operator $\mathcal {M}$ is defined by
\begin{equation}
U(x,t)=\mathcal {M}(u(x,t)).
\end{equation}
For LBM, the operator $\mathcal {M}$ is implemented by
Eq.~(\ref{macrorho}) and Eq.~(\ref{velocity}). Our attention will
put on the development of the reconstruction operator $\mu$ by the
multi-scale analysis. As discussed above the reconstruction operator
in multiscale computation is corresponding to the lifting relation
in an initial problem. In the following we will derive the operator
from the initial problem aspect.

 To obtain an appropriate initial field,
we turn to a simple multiscale perturbation expansion.
We separate the time scale into two different time
scales, $t_1=\epsilon t$ (diffusive time-scale) and $t_2=\epsilon^2
t$ (convective time-scale). The time derivative $\partial_t$ is
expanded using a small parameter $\epsilon$,  which
normally is proportional to the small Knudsen number ($Kn<0.1$)
\cite{ChenDoolen},
\begin{equation}\label{mt}
\partial_t=\epsilon\partial_{t_1}+\epsilon^2\partial_{t_2}+\mathbf{O}(\epsilon^3).
\end{equation}
Similarly, introducing space scale $x_1=\epsilon x$,
the corresponding spatial derivative is not expanded beyond the
first-order term \cite{ChenDoolen}
\begin{equation}\label{malpha}
\partial_\alpha=\epsilon\partial_{1\alpha}+\mathbf{O}(\epsilon^2).
\end{equation}
The single-particle distribution function is expanded as follows
\cite{ChenDoolen}
\begin{equation}\label{mf}
f_i({\rm x},t)=f_i^{(0)}({\rm x},t)+\epsilon f_i^{(1)}({\rm
x},t)+\epsilon^2 f_i^{(2)}({\rm x},t)+\ldots.
\end{equation}
By the Taylor expansion, from Eq. (\ref{lbm}), we get
\begin{equation}\label{taylor}
\delta t(\partial_t+{c}_{i\alpha}\partial_\alpha)f_i({\rm
x},t)+\delta t^2(\partial_t+{c}_{i\alpha}\partial_\alpha)^2f_i({\rm
x},t)+\mathbf{O}(\delta t^3)=\Omega_i.
\end{equation}
Combining Eq.(\ref{mt})-Eq.(\ref{mf}) with Eq.(\ref{taylor}), we
obtain
\begin{equation}
f_i^{(0)}({\rm x},t)=f_i^{\rm(eq)}({\rm x},t)
\end{equation}
and
\begin{equation}\label{expansionchapman}
\begin{array}{c}
\epsilon f_i^{\rm(1)}({\rm x},t)+\epsilon^2 f_i^{(2)}({\rm
x},t)=-\tau_{\rm
lbm}[(\epsilon\partial_{t_1}+\epsilon^2\partial_{t_2}+\epsilon
{c}_{i\alpha}\partial_{1\alpha})\delta
t+\\(\frac{1}{2}\epsilon^2\partial_{t_1}^2{c}_{i\alpha}\partial_{1\alpha}
+\epsilon^2\partial_{t_1}{c}_{i\alpha}\partial_{1\alpha}+\frac{1}{2}\epsilon^2{c}_{i\alpha}{c}_{i\beta}\partial_{1\alpha}\partial_{1\beta})\delta
t^2]\\(f_i^{(0)}({\rm x},t)+\epsilon f_i^{(1)}({\rm
x},t))+\mathbf{O}(\delta t^3).
\end{array}
\end{equation}
For first order of $\epsilon$, we get
\begin{equation}\label{noneq}
f_i^{\rm(1)}({\rm x},t)=-\tau_{\rm lbm}\delta t(\partial_{t_1}+
{c}_{i\alpha}\partial_{1\alpha}({\rm
x},t))f_i^{\rm(eq)}+\mathbf{O}(\delta t^3).
\end{equation}
According to Eq. (\ref{macrorho})-Eq. (\ref{velocity}), we have
following equations in the first-order scale of $\epsilon$
\cite{GuoBook}
\begin{equation}\label{rhot1}
\partial_{t_1}\rho+\partial_{1\alpha}(\rho u_\alpha)+\mathbf{O}(\delta
t^2)=0,
\end{equation}
\begin{equation}\label{velt1}
\partial_{t_1}(\rho u_\alpha)+\partial_{1\beta}(\rho u_\alpha u_\beta+c_s^2\rho\delta_{\alpha\beta})+\mathbf{O}(\delta
t^2)=0,
\end{equation}
Then, Eq. (\ref{velt1}) can be rewritten as
\begin{equation}\label{velt1r}
\rho\partial_{t_1}(u_\alpha)+\rho u_\beta\partial_{1\beta}(\rho
u_\alpha+c_s^2\rho\delta_{\alpha\beta})+\mathbf{O}(\delta t^2)=0.
\end{equation}

By matching small scales, from Eq. (\ref{expansionchapman}), we can
get up to the second order equations of the small parameter
$\epsilon$:
\begin{equation}\label{secondorder}
f_i^{(2)}=-\tau_{\rm lbm}\delta t\partial_{t_2}f_i^{(0)}-\delta
t^2(\tau_{\rm
lbm}-\frac{1}{2})(\partial_{t_1}+c_{i\beta}\partial_{1\beta})^2f_i^{(0)}+\mathbf{O}(\delta
t^3).
\end{equation}
Then, we can get \cite{GuoBook}
\begin{equation}\label{t2rho}
\partial_{t_2}\rho+\mathbf{O}(\delta t^2)=0,
\end{equation}
\begin{equation}\label{t2moment}
\partial_{t_2}(\rho u_\alpha)=\nu\partial_{1\beta}(\rho(\partial_{1\alpha}
u_\beta+\partial_{1\beta} u_\alpha))+\mathbf{O}(\delta t^2+\delta t
u^3)
\end{equation}
 Furthermore, from Eq. (\ref{noneq}), we
have
\begin{equation}\label{noneq_1}
f_i^{(1)}({\rm x},t)=-\tau_{\rm lbm} \delta t(\partial_{t_1}+
{c}_{i\alpha}\partial_{1\alpha})(f_i^{\rm L(eq)}({\rm x},t)+f_i^{\rm
Q(eq)}({\rm x},t))+\mathbf{O}(\delta t^3).
\end{equation}
In the derivation of Eq. (\ref{noneq_1}), we introduce the following
formulas according to the chain rule of derivatives \cite{Imaura}
\begin{equation}
\partial_{t_1}{f_i^{\rm(eq)}({\rm x},t)}=\partial_{\rho}{f_i^{\rm(eq)}({\rm x},t)}\partial_{t_1}{\rho}+
\partial_{u_\beta}{f_i^{(eq)}({\rm x},t)}\partial_{t_1}{u_\beta},
\end{equation}
\begin{equation}
\partial_{1\alpha}{f_i^{\rm(eq)}({\rm x},t)}=\partial_{\rho}{f_i^{\rm(eq)}({\rm x},t)}\partial_{1\alpha}{\rho}+
\partial_{u_\beta}{f_i^{\rm(eq)}({\rm x},t)}\partial_{1\alpha}{u_\beta}.
\end{equation}
Now, the equilibrium function can be differentiated by the
macroscopic variables as follows \cite{Imaura}
\begin{equation}
\partial_{\rho}{f_i^{\rm (eq)}({\rm x},t)}=\frac{1}{\rho}f_i^{\rm(eq)}({\rm x},t),
\end{equation}
\begin{equation}
\partial_{u_\beta}{f_i^{\rm(eq)}({\rm x},t)}=\partial_{u_\beta}{f_i^{\rm L(eq)}({\rm x},t)}+\partial_{u_\beta}{f_i^{\rm Q(eq)}({\rm x},t)}.
\end{equation}
According to Eq. (\ref{leq}) and Eq. (\ref{qeq}), we have
\begin{equation}
\partial_{u_\beta}{f_i^{\rm L(eq)}}=\omega_i\rho\frac{1}{c_s^2}c_{i\beta},
\end{equation}
\begin{equation}
\begin{array}{c}
\partial_{u_\beta}{f_i^{\rm Q(eq)}}=\omega_i\rho\frac{1}{2c_s^4}(
2c_{i\alpha}c_{i\beta}u_{\alpha}-2c_s^2u_{\beta})
=\omega_i\rho(\frac{1}{c_s^4}
c_{i\alpha}c_{i\beta}u_{\alpha}-\frac{1}{c_s^2}u_{\beta}).
\end{array}
\end{equation}
So, we have
\begin{equation}\label{derforeq}
\partial_{u_\beta}{f_i^{\rm(eq)}}=\omega_i\rho[\frac{1}{c_s^2}(c_{i\beta}-u_{\beta})+\frac{1}{c_s^4}
c_{i\alpha}c_{i\beta}u_{\alpha}].
\end{equation}
Come here we can have following corollaries.
\begin{corollary} From Eq. (\ref{noneq_1}), for the first-order approximation of $\epsilon$,
there exists a lifting relation from the macroscopic variables to
the microscopic variable $f_i^{(1)}$
\begin{equation}
\begin{array}{c}
f_i^{(1)}=-\tau_{\rm lbm}\delta
t\{(c_{i\alpha}-u_\alpha)\partial_{1\alpha}\rho\partial_\rho
f_i^{eq}+(c_{i\alpha}-u_\alpha)\partial_{1\alpha}
u_\beta\partial_{u_\beta}f_i^{\rm(eq)}-\\ \rho\partial_\rho
f_i^{\rm(eq)}\partial_\alpha
u_\alpha-\frac{1}{\rho}\partial_{1\alpha}
p\partial_{u_\alpha}f_i^{\rm(eq)}\}\\
=-\tau_{\rm lbm}\delta
t\{(c_{i\alpha}-u_\alpha)\frac{1}{\rho}\partial_{1\alpha}\rho
f_i^{\rm (eq)}+(c_{i\alpha}-u_\alpha)\\ \partial_{1\alpha} u_\beta
\omega_i\rho[\frac{1}{c_s^2}(c_{i\beta}-u_{\beta})+\frac{1}{c_s^4}
c_{i\beta}c_{i\gamma}u_{\gamma}]-\\
f_i^{\rm(eq)}\partial_{1\alpha}
u_\alpha-\frac{1}{\rho}\partial_{1\beta} p\
\omega_i\rho[\frac{1}{c_s^2}(c_{i\beta}-u_{\beta})+\frac{1}{c_s^4}
c_{i\beta}c_{i\gamma}u_{\gamma}]\}.
\end{array}
\end{equation}
\end{corollary}
\begin{corollary} From Eq. (\ref{secondorder}), for the second-order scale of
$\epsilon$, we have the following approximation
\begin{equation}
f_i^{(2)}\approx -\tau_{\rm lbm} \delta
t\partial_{t_2}f_i^{\rm(eq)},
\end{equation}
where the second-order derivative of $f_i^{(0)}$ is ignored.
\end{corollary} Hence,
we can easily establish an approximation for $f_i^{(2)}$ by the
method analogous to the approximation of $f_i^{(1)}$ as follows
\begin{equation}
\partial_{t_2}{f_i^{\rm(eq)}({\rm x},t)}=\partial_{\rho}{f_i^{\rm(eq)}({\rm x},t)}\partial_{t_2}{\rho}+
\partial_{u_\beta}{f_i^{(eq)}({\rm x},t)}\partial_{t_2}{u_\beta}.
\end{equation}
By Eq. (\ref{t2rho}), we have
\begin{equation}
\partial_{t_2}{f_i^{\rm(eq)}({\rm x},t)}=\partial_{u_\beta}{f_i^{\rm(eq)}({\rm x},t)}\partial_{t_2}{u_\beta}
=\frac{1}{\rho}\partial_{u_\beta}{f_i^{\rm(eq)}({\rm
x},t)}\partial_{t_2}{(\rho u_\beta)}.
\end{equation}
From Eq. (\ref{t2moment}) and Eq. (\ref{derforeq}), it is easy to
obtain
\begin{equation}
\partial_{t_2}{f_i^{\rm(eq)}}=\nu \omega_i[\frac{1}{c_s^2}(c_{i\beta}-u_{\beta})+\frac{1}{c_s^4}
c_{i\beta}c_{i\gamma}u_{\gamma}]\partial_{1\alpha}(\rho(\partial_{1\beta}
u_\alpha+\partial_{1\alpha} u_\beta)).
\end{equation}
So, we have
\begin{equation}\label{50}
\epsilon^2f_i^{(2)}\approx-\tau\delta t\nu
\omega_i[\frac{1}{c_s^2}(c_{i\beta}-u_{\beta})+\frac{1}{c_s^4}
c_{i\beta}c_{i\gamma}u_{\gamma}]\partial_{\alpha}(\rho(\partial_{\beta}
u_\alpha+\partial_{\alpha} u_\beta)).
\end{equation}
By a simple derivation, we have
\begin{equation}\label{51}
\partial_{\alpha}(\rho(\partial_{\beta}
u_\alpha+\partial_{\alpha}
u_\beta))=\partial_{\alpha}\rho(\partial_{\beta}
u_\alpha+\partial_{\alpha}
u_\beta)+\rho(\partial_{\beta}\partial_{\alpha}
u_\alpha+\partial_{\alpha}^2 u_\beta).
\end{equation}
From Eqs. (\ref{50})$\sim$(\ref{51}), we have
\begin{equation}
\begin{array}{c}
\epsilon^2f_i^{(2)}\approx-\tau_{\rm lbm}\delta t\nu
\omega_i[\frac{1}{c_s^2}(c_{i\beta}-u_{\beta})+\frac{1}{c_s^4}
c_{i\beta}c_{i\gamma}u_{\gamma}](\partial_{\alpha}\rho(\partial_{\beta}
u_\alpha+\partial_{\alpha} u_\beta)+\\
\rho(\partial_{\beta}\partial_{\alpha} u_\alpha+\partial_{\alpha}^2
u_\beta))
\end{array}
\end{equation}

 Therefore, we get the following approximation of the
non-equilibrium distribution function from Eq. (\ref{mf})
\begin{equation}\label{53}
f_i^{\rm(neq)}\approx \epsilon f_i^{(1)}+\epsilon^2 f_i^{(2)},
\end{equation}
that is,
\begin{equation}
\begin{array}{c}
f_i^{\rm(neq)}({\rm x},t)\approx -\tau_{\rm lbm}\delta
t\{u_{T,i\alpha}\frac{1}{\rho}\partial_{\alpha}\rho f_i^{\rm
(eq)}+u_{T,i\alpha}\partial_{\alpha} u_\beta
\omega_i\rho[\frac{1}{c_s^2}u_{T,i\beta}+\frac{1}{c_s^4}
c_{i\beta}c_{i\gamma}u_{\gamma}]\\ - f_i^{\rm(eq)}\partial_{\alpha}
u_\alpha-\frac{1}{\rho}\partial_{\beta} p\
\omega_i\rho[\frac{1}{c_s^2}u_{T,i\beta}+\frac{1}{c_s^4}
c_{i\beta}c_{i\gamma}u_{\gamma}]\\
+\omega_i\rho[\frac{1}{c_s^2}u_{T,i\beta}+\frac{1}{c_s^4}
c_{i\beta}c_{i\gamma}u_{\gamma}](\nu\frac{1}{\rho}\partial_{\alpha}\rho(\partial_{\beta}
u_\alpha+\partial_{\alpha} u_\beta)+
\nu(\partial_{\beta}\partial_{\alpha} u_\alpha+\partial_{\alpha}^2
u_\beta))\},
\end{array}
\end{equation}
where $u_{T,i\alpha}=c_{i\alpha}-u_\alpha$ (${\rm u}_T={\rm
c}_i-{\rm u}$, peculiar velocity). Since the velocity
field is divergence-free, we have
\begin{equation}
\begin{array}{c}
f_i^{\rm(neq\_dfv)}\approx -\tau_{\rm lbm}\delta
t\{u_{T,i\alpha}\frac{1}{\rho}\partial_{\alpha}\rho f_i^{\rm
(eq)}+u_{T,i\alpha}\partial_{\alpha} u_\beta
\omega_i\rho[\frac{1}{c_s^2}u_{T,i\beta}+\frac{1}{c_s^4}
c_{i\beta}c_{i\gamma}u_{\gamma}]-\\
-\frac{1}{\rho}\partial_{1\beta} p\
\omega_i\rho[\frac{1}{c_s^2}u_{T,i\beta}+\frac{1}{c_s^4}
c_{i\beta}c_{i\gamma}u_{\gamma}]\\
+\omega_i\rho[\frac{1}{c_s^2}u_{T,i\beta}+\frac{1}{c_s^4}
c_{i\beta}c_{i\gamma}u_{\gamma}](\nu\frac{1}{\rho}\partial_{\alpha}\rho(\partial_{\beta}
u_\alpha+\partial_{\alpha} u_\beta)+ \nu\partial_{\alpha}^2
u_\beta)\}.
\end{array}
\end{equation}
 Here, we also introduce an
approximation of $\partial_{u_\beta} f_i^{\rm(eq)}$ by ignoring the
higher-order terms of $u^2$ as adopted in \cite{Imaura}
\begin{equation}
\partial_{u_\beta} f_i^{\rm(eq)}=\omega_i\rho[\frac{1}{c_s^2}u_{T,i\beta}+\frac{1}{c_s^4}
c_{i\beta}c_{i\gamma}u_{\gamma}]\approx\frac{u_{T,i\beta}}{c_s^2}f_i^{\rm
(eq)}.
\end{equation}
Now, we have
\begin{equation}
\begin{array}{c}
f_i^{\rm(neq\_dfv)}\approx  -\tau_{\rm lbm}\delta
t\frac{1}{c_s^2}f_i^{\rm (eq)}u_{T,i\beta}
(u_{T,i\alpha}\partial_{\alpha}u_\beta+
\nu\frac{1}{\rho}\partial_{\alpha}\rho(\partial_{\beta}
u_\alpha+\partial_{\alpha} u_\beta)+\nu\partial_{\alpha}^2 u_\beta).
\end{array}
\end{equation}
Rewriting the above formula, we obtain
\begin{equation}\label{fneqdfv}
\begin{array}{c}
f_i^{\rm(neq\_dfv)}\approx-\tau_{\rm lbm}\delta t
f_i^{\rm(eq)}\frac{1}{c_s^2}u_{T,i\beta}(u_{T,i\alpha}\partial_{\alpha}u_\beta+\nu
\partial_{\alpha}^2 u_\beta+ \nu\frac{1}{\rho}\partial_{\alpha}\rho S_{\alpha\beta}),
\end{array}
\end{equation}
where $S_{\alpha\beta}=\partial_{\beta} u_\alpha+\partial_{\alpha}
u_\beta$.

In all, we can get an approximation of the single-particle
distribution function for divergence-free velocity fields as follows
\begin{equation}\label{distrib}
\begin{array}{c}
f_i\approx f_i^{\rm(eq)}\{1-\frac{1}{c_s^2}\tau_{\rm lbm}\delta t
u_{T,i\beta}(u_{T,i\alpha}\partial_{\alpha}u_\beta+\nu
\partial_{\alpha}^2 u_\beta+ \nu\frac{1}{\rho}\partial_{\alpha}\rho S_{\alpha\beta})\},
\end{array}
\end{equation}

By a similar deviation, we can get an approximation of the
single-particle distribution function for weak-compressible velocity
fields as follows:
\begin{equation}\label{distribdiv}
\begin{array}{c}
f_i\approx f_i^{\rm(eq)}\{1-\frac{1}{c_s^2}\tau_{\rm lbm}\delta t
[u_{T,i\beta}(u_{T,i\alpha}\partial_{\alpha}u_\beta+\nu(
\partial_{\alpha}^2 u_\beta+\partial_{\beta}\partial_{\alpha} u_\alpha)\\+ \nu\frac{1}{\rho}\partial_{\alpha}\rho S_{\alpha\beta})-c_s^2\partial_\alpha u_\alpha]\},
\end{array}
\end{equation}

Now we compare our results with that published in literatures.

1. T. Imamura {\it et al}\cite{Imaura} obtained the following
formula
\begin{equation}\label{Imamuraf}
{f}_i\approx
{f}^{\rm(eq)}_i+\epsilon{f}^{\rm(1)}_i={f}^{\rm(eq)}_i[1-\tau_{\rm
lbm}\delta t(\frac{3
u_{T,i\alpha}u_{T,i\beta}}{c^2}-\delta_{\alpha\beta})\partial_\beta
u_\alpha]
\end{equation}
They only used ${f}^{\rm(1)}_i$ to approximate the single-partial
distribution functions. It is well-known that in order to recover
the correct Navier-Stokes equations, ${f}^{\rm(2)}_i$ is needed.
From this point of view, the approaching form (\ref{distribdiv}) of
the distribution functions are more accurate than (\ref{Imamuraf}).
If the divergence-free velocity field is considered, Eq.
(\ref{distrib}) is also superior to Eq. (\ref{Imamuraf}) because Eq.
(\ref{distrib}) contains the information of ${f}^{(2)}_i$ which is
related with molecule viscosity and density gradient. As for the
lifting relation it is certainly essential to involve molecule
viscosity and density gradient \cite{Chopard,Leemput}.

2. Skodors \cite{Skordos} gave the following formula (ignoring the
term of $O(Ma^2)$)
\begin{equation}\label{skodorsneq}
f_i^{\rm (neq),S}=-\tau_{\rm
lbm}\delta_t\omega_i\left[\frac{1}{c_s^2}{\rm c}_i{\rm
c}_i:\nabla(\rho{\rm u})-\nabla\cdot(\rho{\rm u})\right].
\end{equation}
Guo and Zhao \cite{guo} further simplified Eq.(\ref{skodorsneq}) and
obtained the following relation
\begin{equation}\label{Guo}
f_i^{\rm (neq),G}=-\tau_{\rm
lbm}\delta_t\omega_i\frac{\rho_0}{c_s^2}{\rm c}_i{\rm
c}_i:\nabla{\rm u}.
\end{equation}
It is very clear that Eqs.(\ref{distrib}) and (\ref{distribdiv}) are
totally different from Eqs.(\ref{skodorsneq}) and (\ref{Guo}),
respectively. The co-existence of Eqs.
(\ref{distrib})$\sim$(\ref{distribdiv}) and
Eqs.(\ref{skodorsneq})$\sim$(\ref{Guo}) as the lift correlation for
the same situation may be regarded as the witness that the
transformation from one-to-many must necessarily be not unique
\cite{NIE}. Some comparisons will be performed in Sec. 4 between
Eqs. (\ref{distrib})$\sim$(\ref{distribdiv}) and
Eqs.(\ref{skodorsneq})$\sim$\ref{Guo}) for schemes of D2Q9 and
D2Q17. It turns out that the accuracy of Eqs.
(\ref{distrib})$\sim$(\ref{distribdiv}) derived in this paper is
better than that of Eqs.(\ref{skodorsneq})$\sim$(\ref{Guo}). The
derivation procedures of Eqs.
(\ref{distrib})$\sim$(\ref{distribdiv}) kept the information of the
$f^{\rm (2)}_2$ and other more details which are important to reduce
the reconstruction relative errors.

\section{Derivation of Non-equilibrium distributions via Boltzmann-BGK equations}\label{IS}

The Boltzmann equation \cite{Chapman} describes the statistical
distribution of particles in a fluid. It is one of the most
important equations of non-equilibrium statistical mechanics, which
deals with systems far from thermodynamic equilibrium
\cite{Schwabl}. The Boltzmann equation is described by
\begin{equation}\label{Beq}
\frac{\partial f({\rm x},{\rm v},t)}{\partial t}+{\rm
v}\cdot\nabla_x f({\rm x},{\rm v},t)+\frac{1}{m}F(x)\nabla_v f({\rm
x},{\rm v},t)=\Omega(f({\rm x},{\rm v},t)).
\end{equation}
The Boltzmann equation (\ref{Beq}) is an equation for the time $t$
evolution of the distribution (properly a density) function $f({\rm
x},{\rm v},t)$ in one-particle phase space, where ${\rm
x}=(x_1,x_2,\cdots,x_d)\in R^{\rm d}$ and ${\rm
v}=(v_1,v_2,\cdots,v_d)\in R^{\rm d}$ (${\rm d}$ denotes the spatial
dimension) are position and velocity, respectively. The equilibrium
distribution function $f^{\rm eq}({\rm x},{\rm v},t)$ can be
determined by
\begin{equation}\label{Maxwell}
f^{(\rm eq)}({\rm x},{\rm v},t)=n({\rm x},t)\left(\frac{m}{2\pi
\kappa T({\rm x},t)}\right)^{d/2}{\rm exp}\left[-\frac{m}{2\kappa
T({\rm x},t)}({\rm v}-{\rm u}({\rm x},t))^2\right]
\end{equation}
Here, the quantities $T({\rm x},t)$, $n({\rm x},t)$ and ${\rm
u}({\rm x},t)$ represent the \textit{local temperature}, the
\textit{local particle-number distribution density} and the
\textit{local velocity} \cite{Succi,Schwabl}, repectively. ${\rm
u_T}={\rm v}-{\rm u}({\rm x},t)$ is the so called thermal velocity.
$m$ denotes the single-particle mass which is set to be unity for
convenience. In order to simplify the complex collisional term, the
following \textit{conserved relaxation time approximation} is used
to describe the collision term through only one characteristic
frequency\cite{Schwabl}
\begin{equation}\label{lBeq}
\frac{\partial f({\rm x},{\rm v},t)}{\partial t}+{\rm v}\cdot\nabla
f({\rm x},{\rm v},t)=-\frac{1}{\tau}(f({\rm x},{\rm v},t)-f^{(\rm
eq)}({\rm x},{\rm v},t)),
\end{equation}
where the external force term is not considered and $\nabla$ denotes
$\nabla_x$. $\tau$ represents the relaxation time.

 In order to solve Eq.(\ref{lBeq}), the velocity space is discretized
 \cite{Succi} and we gain
\begin{equation}\label{dlBeq}
\frac{\partial f_i({\rm x},t)}{\partial t}+{{\rm c}_i}\cdot\nabla
f_i({\rm x},t)=-\frac{1}{\tau}(f_i({\rm x},t)-f_i^{(\rm eq)}({\rm
x},t)),
\end{equation}
where $w_i$ denotes the integral weight factor, $f_i({\rm x},t)=w_i
f({\rm x},{\rm c}_i,t)$ and $f_i^{(\rm eq)}({\rm x},t)=w_i f^{(\rm
eq)}({\rm x},{\rm c}_i,t)$. Furthermore, along the characteristic
line, the time-discretization form of Eq.(\ref{dlBeq}) can be
expressed as \cite{Succi,Qian}
\begin{equation}\label{lbm2}
f_i({\rm x}+{\rm c}_i\delta t,t+\delta t)=f_i({\rm
x},t)-\frac{1}{\tau_{\rm lbm}}(f_i({\rm x},t)-f_i^{(\rm eq)}({\rm
x},t)),i=0,1,...,N.
\end{equation}
where $f_i$ is the probability distribution function (PDF) along the
ith direction, $f_i^{(\rm eq)}$ is its corresponding equilibrium
PDF, $\delta t$ is the time step, ${\rm c}_i$ is the particle
velocity in the ith direction, and $N$ is the number of the discrete
particle velocities. Note: $\tau_{\rm lbm}=\tau/\delta t$ which is a
dimensionless relaxation time. The local macro quantities are
defined by Eqs. (\ref{macrorho}) and (\ref{velocity}).

At the low fluid flow velocity (or low Mach number), an approximate
form of the equilibrium distribution function $f_i^{(\rm eq)}$ is
described by the discrete equilibrium distribution, Eqs.
(\ref{equilibrium})$\sim$(\ref{qeq}).

 Now, we consider the \textit{conserved relaxation
time approximation} of the Boltzmann equation (\ref{lBeq}). The
right hand side of Eq. (\ref{lBeq}) represents the difference
between the distribution function and a local Maxwell distribution.
This difference is termed non-equilibrium distribution defined by
\begin{equation}
f^{(\rm neq)}({\rm x},{\rm v},t)=f({\rm x},{\rm v},t)-f^{(\rm
eq)}({\rm x},{\rm v},t).
\end{equation}
Then, Eq. (\ref{lBeq}) can be rewritten as follows
\begin{equation}\label{divideneq}
\left(\frac{\partial}{\partial t}+{\rm v}\cdot \nabla\right)f^{(\rm
neq)}({\rm x},{\rm v},t)+\left(\frac{\partial}{\partial t}+{\rm
v}\cdot \nabla\right)f^{(\rm eq)}({\rm x},{\rm
v},t)=-\frac{1}{\tau}f^{(\rm neq)}({\rm x},{\rm v},t).
\end{equation}
In the hydrodynamic region \cite{Chapman}, the first term on the
left-hand side of Eq. (\ref{divideneq}) can be neglected compared
with the right-hand side \cite{Schwabl}. Then, we obtain
\begin{equation}\label{sdivideneq}
\left(\frac{\partial}{\partial t}+{\rm v}\cdot \nabla\right)f^{(\rm
eq)}({\rm x},{\rm v},t)=-\frac{1}{\tau}f^{(\rm neq)}({\rm x},{\rm
v},t).
\end{equation}
In terms of the Maxwell equilibrium distribution and assuming a
uniform temperature of the system, we can obtain
\begin{equation}\label{substituteEq}
\begin{array}{c}
\frac{f^{(\rm eq)}({\rm x},{\rm v},t)}{n({\rm
x},t)}\left(\frac{\partial}{\partial t}+{\rm v}\cdot
\nabla\right)n({\rm x},t)-f^{(\rm eq)}({\rm x},{\rm
v},t)\left(\frac{\partial}{\partial t}+{\rm v}\cdot
\nabla\right)\frac{{\rm u_{T}^2}}{2\kappa
T}\\=-\frac{1}{\tau}f^{(\rm neq)}({\rm x},{\rm v},t),
\end{array}
\end{equation}
where $T=T({\rm x},t)={\rm constant}$. In Eq. (\ref{substituteEq}),
the left-hand term can be rewritten as follows
\begin{equation}\label{masscon}
\begin{array}{c}
\frac{f^{(\rm eq)}({\rm x},{\rm v},t)}{n({\rm
x},t)}\left(\frac{\partial}{\partial t}+{\rm v}\cdot
\nabla\right)n({\rm x},t)=\frac{f^{(\rm eq)}({\rm x},{\rm
v},t)}{n({\rm x},t)}\left(\frac{\partial}{\partial t}+{\rm u({\rm
x},{\rm t})}\cdot \nabla\right)n({\rm x},t)\\+\frac{f^{(\rm
eq)}({\rm x},{\rm v},t)}{n({\rm x},t)}{\rm u_{\rm T}}\cdot \nabla
n({\rm x},t)
\end{array}
\end{equation}
In order to satisfy the mass conservation condition of the fluid
flow system, the first term of the right-hand side in Eq.
(\ref{masscon}) should be equal to zero. Hence, we have the
following equation
\begin{equation}\label{thermalvelocity}
\begin{array}{c}
f^{(\rm eq)}({\rm x},{\rm v},t)\left(\frac{\partial}{\partial
t}+{\rm v}\cdot \nabla\right)\frac{{\rm
u_{T}^2}}{2c_s^2}-\frac{f^{(\rm eq)}({\rm x},{\rm v},t)}{n({\rm
x},t)}({\rm u_{\rm T}}\cdot \nabla n({\rm x},t)-n({\rm
x},t)\nabla\cdot {\rm u}({\rm x},t))=\\ \frac{1}{\tau}f^{(\rm
neq)}({\rm x},{\rm v},t), \end{array}
\end{equation}
where $c_s=\sqrt{\kappa T}$. The term ${\rm u_T^2}$ is the thermal
fluctuation energy, thus the non-equilibrium is determined by the
material derivative of this thermal fluctuation energy. The quantity
$\left(\frac{\partial}{\partial t}+{\rm v}\cdot \nabla\right){\rm
u_T^2}$ can be determined by the dynamical equation corresponding to
the micro dynamical system. Here, we rewrite
$\frac{1}{2}\left(\frac{\partial}{\partial t}+{\rm v}\cdot
\nabla\right){\rm u_T^2}$ as follows
\begin{equation}\label{77}
\frac{1}{2}\left(\frac{\partial}{\partial t}+{\rm v}\cdot
\nabla\right){\rm u_T^2}=-{\rm
u_T}\cdot\left(\frac{\partial}{\partial t}+{\rm u({\rm x},{\it
t})}\cdot \nabla\right){\rm u}({\rm x},t)-{\rm u_T}\cdot({\rm
u_T}\cdot\nabla){\rm u}({\rm x},t).
\end{equation}
Generally, the governing equation of the macroscopic physical
quantity is represented by
\begin{equation}\label{force}
\frac{D}{Dt}{\rm u}({\rm x},t)=\left(\frac{\partial}{\partial
t}+{\rm u({\rm x},{\it t})}\cdot \nabla\right){\rm u}({\rm
x},t)=F({\rm x},{\rm u(x,{\it t})},t).
\end{equation}
Normally, the macroscopic physical quantity ${\rm u}({\rm x},t)$ in
the governing equation is known. So, $F({\rm x}, {\rm u(x,{\it
t})},t)$ can be determined easily. For fluid flow problems, taking
${\rm u}({\rm x},t)$ as fluid velocity, then $F({\rm x},{\rm
u(x,{\it t})},t)$ can be estimated by fluid acceleration. The term
${\rm u_T}\cdot({\rm u_T}\cdot\nabla){\rm u}$ in Eq. (\ref{77}) can
be determined by ${\rm u(x,{\it t})}$ and the spatial derivatives of
${\rm u(x,{\it t})}$.

The lattice Boltzmann model is a special discrete
form of the BGK lattice Bolzmann equation with respect to temporal
and spatial variables. For LBM the equilibrium distribution, Eq.
(\ref{equilibrium}), is a polynomial-truncated approximation of the
Maxwell distribution up to $O(|{\rm u}|^3)$, so Eq.
(\ref{thermalvelocity}) can be applied to LBM directly as follows
\begin{equation}
\begin{array}{c}
f_i^{(\rm eq)}({\rm x},t)\left(\frac{\partial}{\partial t}+{\rm
c}_i\cdot \nabla\right)\frac{{\rm u}_{i,{\rm
T}}^2}{2c_s^2}-\frac{f^{(\rm eq)}_{i}({\rm x},{\rm v},t)}{n({\rm
x},t)}({\rm u}_{i,{\rm T}}\cdot \nabla n({\rm x},t)-n({\rm
x},t)\nabla\cdot {\rm u}({\rm x},t))=\\ \frac{1}{\tau}f_i^{(\rm
neq)}({\rm x},t), \end{array}
\end{equation}
where ${{\rm u}_{i,{\rm T}}={\rm c}_i-{\rm u}({\rm x},{\it t})}$.
Now, the non-equilibrium distribution function can be denoted by
\begin{equation}\label{lbmthermalvelocity}
\begin{array}{r}
f_i^{(\rm neq)}({\rm x},t)=-\frac{\tau f_i^{(\rm eq)}({\rm
x},t)}{c_s^2}\left[{\rm u}_{i,{\rm T}}\cdot \left(F({\rm x},{\rm
u(x,{\it t})},t)+({\rm u}_{i,{\rm T}}\cdot\nabla){\rm
u}+\frac{c_s^2}{n({\rm x}, t)}\nabla n({\rm
x},t)\right)-c_s^2\nabla\cdot {\rm u}({\rm x},t)\right].
\end{array}
\end{equation}

The derivation of Eq.(\ref{lbmthermalvelocity}) is completed based
on the rigorous inherent physical consistency in the hydrodynamic
region and and the derivation is independent on the spatial
dimension. Meanwhile, the Maxwell equilibrium distribution is
regarded as the tool to implement the analysis.

It is worth pointing out that for DnQb LBM, $F({\rm x},{\rm u(x,{\it
t})},t)$ can easily be  determined from the recovered Naiver-Stokes
equations, so the obtained non-equilibrium distribution function
formulas  (\ref{lbmthermalvelocity}) and (\ref{distribdiv}) are
identical. Thus, by using different derivation method we come to the
same conclusion.

In addition, according to Eqs. (\ref{distrib}),(\ref{distribdiv})
and (\ref{lbmthermalvelocity}), it can be seen that the
non-equilibrium distribution functions have the following form
\begin{equation}\label{pert}
f_i^{(\rm neq)}=f_i^{(\rm eq)}\lambda_i(\rho,{\rm u}),
\end{equation}
where $\lambda_i(\rho,{\rm u})$ is a perturbative parameter with
respect to $\rho$ and ${\rm u}$. The parameter $\lambda_i(\rho,{\rm
u})$ in Eq. (\ref{pert}) needs to satisfy the following constraints
\begin{equation}
\sum_i f_i^{(\rm eq)}\lambda_i(\rho,{\rm u})=0,\ \sum_{{\rm c}_i\in
\mathcal {V}} {\rm c}_if_i^{(\rm eq)}\lambda_i(\rho,{\rm u})=0
\end{equation}

\section{Numerical Tests}\label{NT}
In this section, the non-equilibrium distribution function will be
validated by numerical tests. The numerical tests focus on
validating the precision of the reconstruction operator and the
correctness of the coupling computations. It's worth
noting that the word ``multiscale simulation" used in this paper is
referred to the coupling between numerical methods of microscale
(molecular dynamics simulation), mesoscale (LBM) and macroscale(say,
FVM) adopted in neighboring computational regions. And for such
coupling the major concern is the transformation of solutions from
macro(or meso)scales to meso(or micro)scales at the interface. The
focus of the following presentation is to validate the correctness
of the proposed operators. Because of space limitation the effect of
the grid fineness on the numerical solution will not be conducted.
Reference \cite{LuanXu} can be referred. The effect of the mesh size on the accuracy of the reconstruction operator will be presented in Sec. \ref{convergence}.

\subsection{Examination of the precision of the reconstruction operator}\label{turb}
 In order to validate Formula
(\ref{lbmthermalvelocity}), the D2Q9 \cite{Qian} and D2Q17
\cite{qian2} LBM are adopted to simulate 2D fluid flows. At low Mach
number ($ Ma=\rm u(x,{\it t})/c_s\ll 1$), the R.H.S of
Eq.(\ref{force}) is equal to the R.H.S of Eq. (\ref{finalNSE})
\begin{eqnarray}\label{finalNSE2}
F_{\alpha}({\rm x},{\rm u(x,{\it t})},t)=-\frac{\partial_\alpha
p}{\rho}+\nu(\partial_\beta\partial_\beta
 u_\alpha+\partial_\alpha\partial_\beta u_\beta)+\nu\frac{\partial_\beta \rho}{\rho}(\partial_\alpha u_\beta+\partial_\beta
 u_\alpha)
\end{eqnarray}
where
\begin{equation}
\nu=c_s^2(\tau_{\rm lbm}-\frac{1}{2})\delta t.
\end{equation}
 The details of the macroscopic dynamic equation corresponding
to D2Q17 LBM are omitted (see \cite{qian2}). Now, the
non-equilibrium distribution in Eq.(\ref{lbmthermalvelocity}) can be
determined directly by the right-hand side of Eq.(\ref{finalNSE}).
For any given initial velocity and density fields, each term in the
right-hand side of Eq.(\ref{finalNSE}) can be calculated. In order
to validate the precision of the proposed method, the following two
basic quantities are defined
\begin{equation}
\widehat{f}_i({\rm x},t)=f_i^{(\rm eq)}({\rm
x},t)+\widehat{f}_i^{(\rm neq)}({\rm x},t),
\end{equation}
\begin{equation}
f_i^{(\rm neq)}({\rm x},t)=f_i({\rm x},t)-f_i^{(\rm eq)}({\rm x},t)
\end{equation}
where $\widehat{f}_i^{(\rm neq)}({\rm x},t)$ is called reconstructed
non-equilibrium distribution function and is calculated by
Eq.(\ref{lbmthermalvelocity}) and $\widehat{f}_i({\rm x},t)$ is the
reconstructed single-particle distribution function. $f_i^{(\rm
neq)}({\rm x},t)$ and $f_i({\rm x},t)$ denote the real
non-equilibrium distribution function and the real single-particle
distribution function, respectively. Here, we give two kinds of relative error definitions: single particle
distribution function reconstruction error, single particle
non-equilibrium distribution function reconstruction error
\begin{equation}
{\rm E}(f_i,\widehat{f}_i)=\sqrt{\frac{1}{Num\times (n+1)}\sum_x
\sum_i\frac{|\widehat{f}_i({\rm x},t)-f_i({\rm x},t)|^2}{f_i({\rm
x},t)^2}},
\end{equation}
\begin{equation}
{\rm
E}(f_i^{\rm(neq)},\widehat{f}_i^{\rm(neq)})=\sqrt{\frac{1}{Num\times(
n+1)}\sum_x \sum_i\frac{|\widehat{f}_i^{\rm(neq)}({\rm
x},t)-f_i^{\rm(neq)}({\rm x},t)|^2}{f_i^{\rm(neq)}({\rm x},t)^2}}
\end{equation}
where $Num$ denotes the number of lattice nodes. 

 In order to demonstrate the proposed method, a freely-decaying 2D
turbulence problem will be simulated by the proposed method. This
turbulence problem often makes the local discrete single-particle
distribution functions to be far from the local discrete equilibrium
distribution functions, which yields a rich velocity structure. The
freely-decaying 2D turbulence is implemented in a periodic box
$\Omega=[0,2\pi]\times[0,2\pi]$. A 2D random velocity field will be
specified as the initial condition. The initial fields are
initialized by random phase in Fourier spectral space and the
initial spectrum is given by \cite{Chasnov}
\begin{equation}
E(k,0)=a_su_0^2k_p^{-1}\left(\frac{k}{k_p}\right)^{(2s+1)}{\rm
exp}\left[-\left(s+\frac{1}{2}\right)\left(\frac{k}{k_p}\right)^2\right]
\end{equation}
where $s=0,1,2,\cdots,$ and the normalization constant $a_s$ is
given by
$$
a_s=(2s+1)^{s+1}/2^ss!.
$$
All the results presented below correspond to $s=3$, $k_p=16$,
$u_0=\{0.1, 0.01\}$ and $\rho=2.7$. The lattice size is $512\times
512$. The integral length scale $L$ is equal to 0.12953. The
Reynolds number ($Re_L=Lu_0/\nu$) is equal to $111.4$.

\begin{figure}
\center{\includegraphics{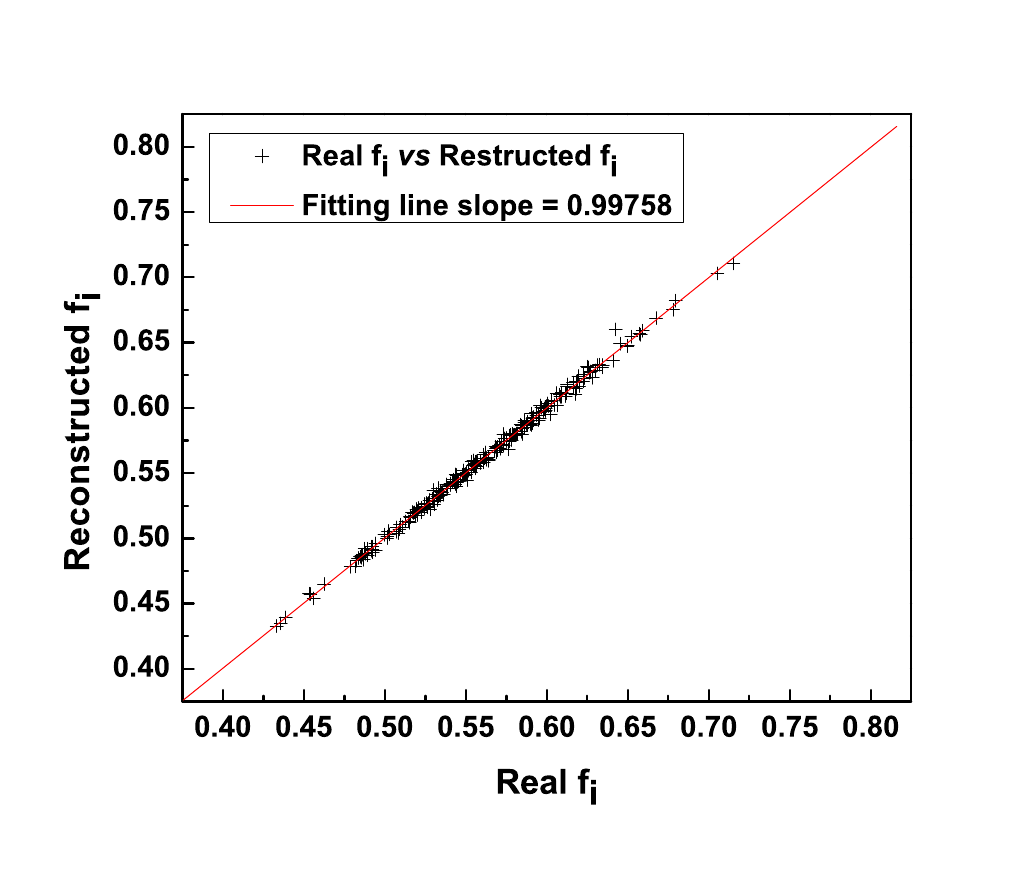}}\center{(a)Linear
regression between $f_i({\rm x},t)$ and $\widehat{f}_i({\rm x},t)$ }
\center{\includegraphics{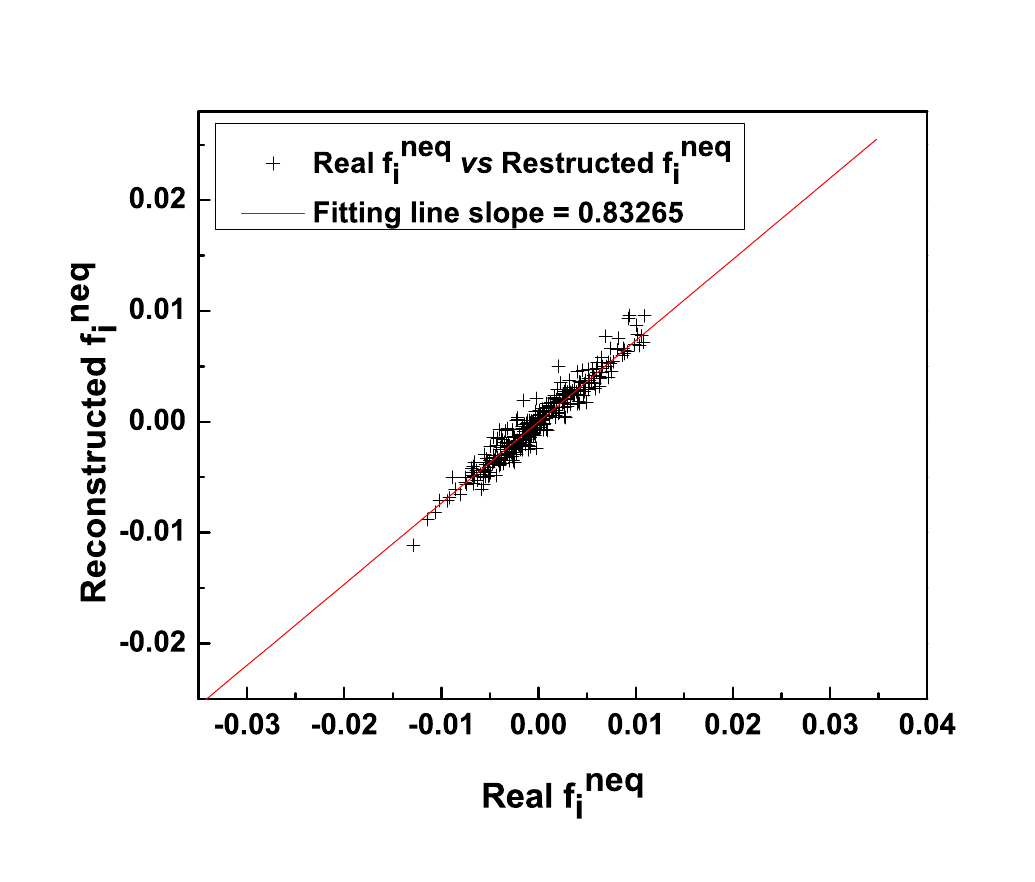}}\center{(b)Linear
regression between $f_i^{(\rm neq)}({\rm x},t)$ and
$\widehat{f}_i^{(\rm neq)}({\rm x},t)$}\caption{Linear regression
(D2Q9, $u_0=0.1$, $t=1000\delta t$, $i=2$): (a)Fit the line
$\widehat{f}_i({\rm x},t)=af_i({\rm x},t)+b$,where $a=0.99758$ and
$b=0.00135$;(b)Fit the line $\widehat{f}_i^{(\rm neq)}({\rm
x},t)=af_i^{(\rm neq)}({\rm x},t)+b$,where $a=0.83265$ and
$b=-2.95012\times 10^{-6}$. Standard
deviation:(a)$\sigma=0.00308$;(b)$\sigma=9.21597\times 10^{-4}$.}
\label{fig.1}
\end{figure}

\begin{figure}
\center{\includegraphics{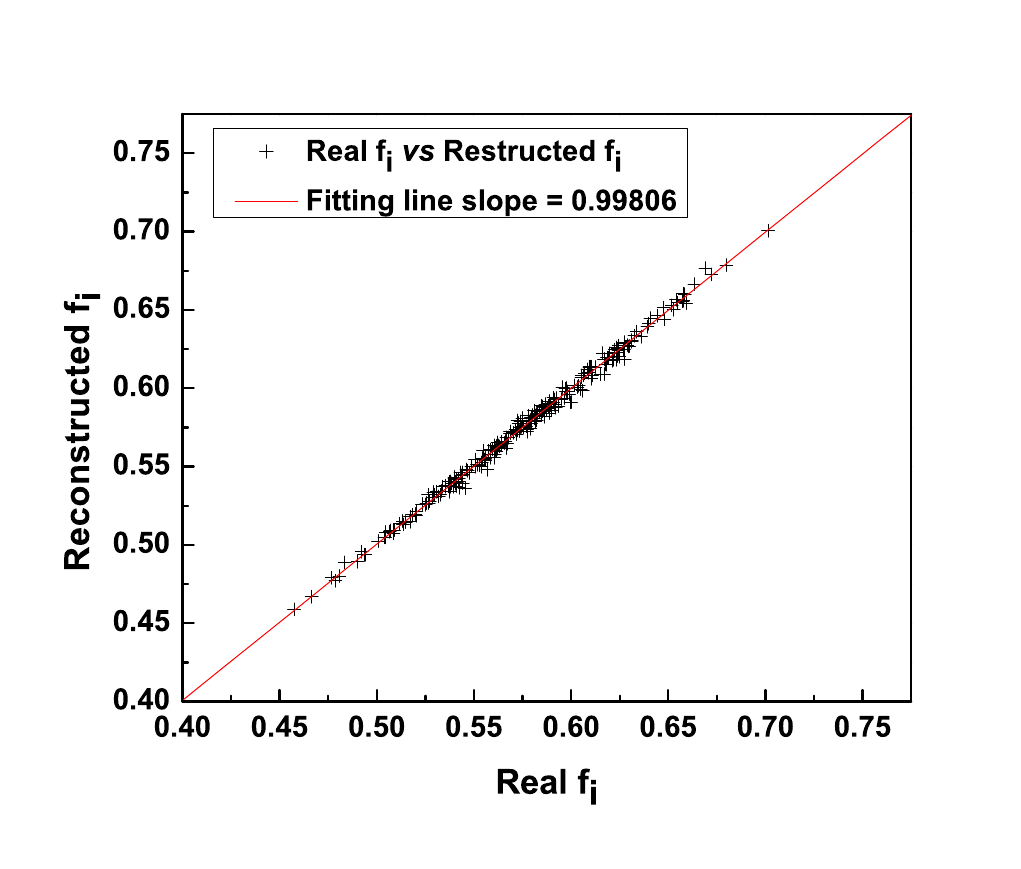}}\center{(a)Linear
regression between $f_i({\rm x},t)$ and $\widehat{f}_i({\rm x},t)$ }
\center{\includegraphics{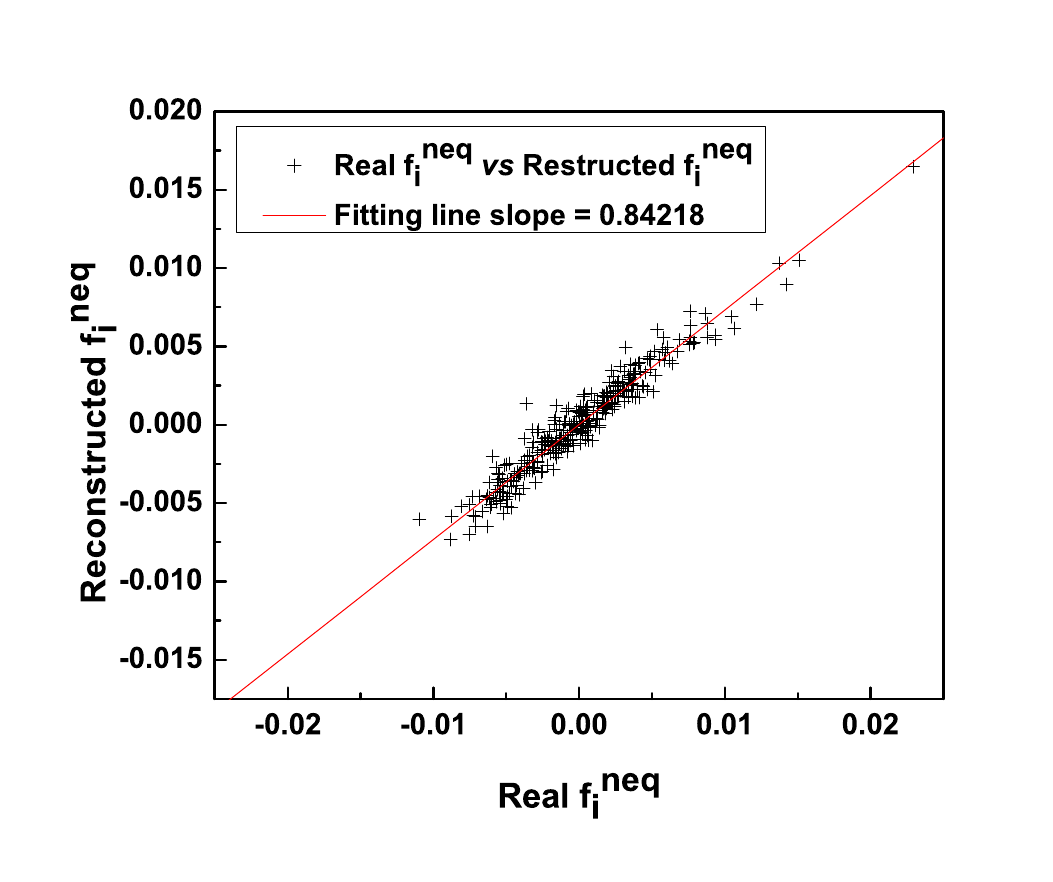}}\center{(b)Linear
regression between $f_i^{(\rm neq)}({\rm x},t)$ and
$\widehat{f}_i^{(\rm neq)}({\rm x},t)$}\caption{Linear regression
(D2Q17, $u_0=0.1$ $t=1000\delta t$, $i=2$): (a)Fit the line
$\widehat{f}_i({\rm x},t)=af_i({\rm x},t)+b$,where $a=0.99806$ and
$b=0.00227$;(b)Fit the line $\widehat{f}_i^{(\rm neq)}({\rm
x},t)=af_i^{(\rm neq)}({\rm x},t)+b$,where $a=0.84218$ and
$b=-4.84408\times 10^{-6}$. Standard
deviation:(a)$\sigma=0.00288$;(b)$\sigma=8.39673\times 10^{-4}$.}
\label{fig.2}
\end{figure}

\begin{figure}
\center{\includegraphics{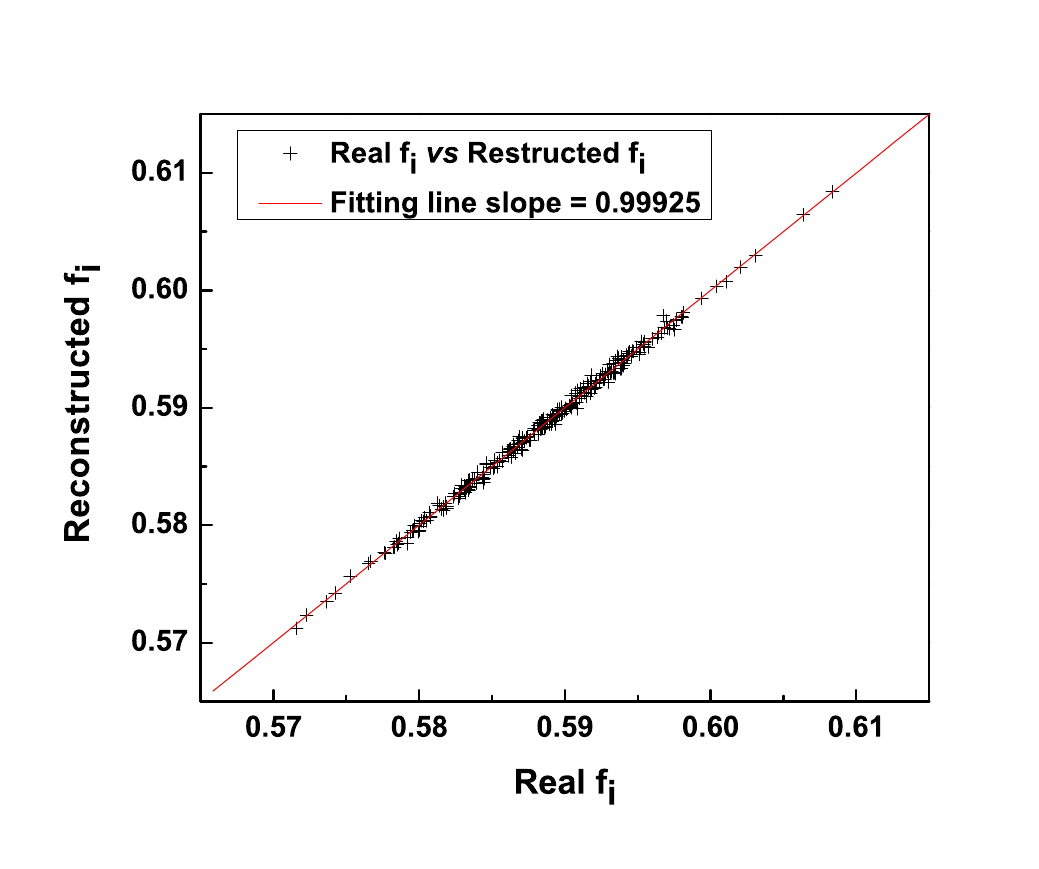}}\center{(a)(a)Linear
regression between $f_i({\rm x},t)$ and $\widehat{f}_i({\rm x},t)$}
\center{\includegraphics{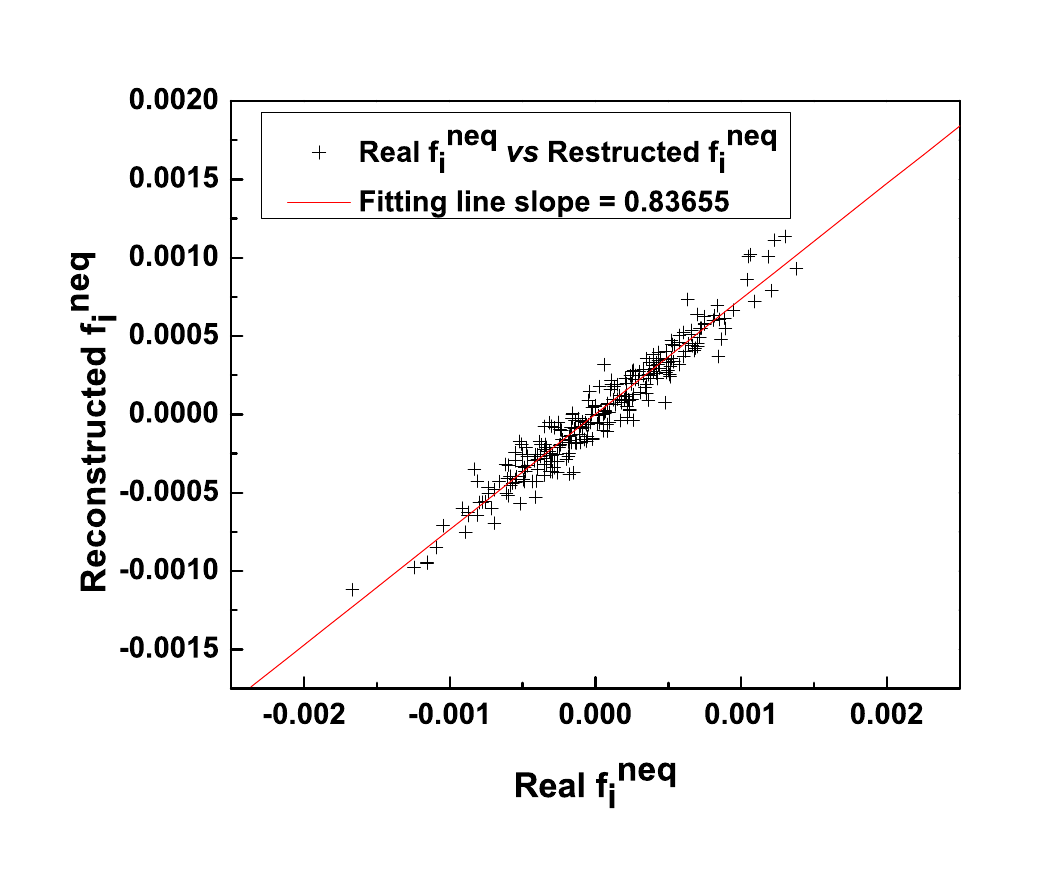}}\center{(b)Linear
regression between $f_i^{(\rm neq)}({\rm x},t)$}\caption{Linear
regression (D2Q9, $u_0=0.01$, $t=10000\delta t$, $i=2$): (a)Fit the
line $\widehat{f}_i({\rm x},t)=af_i({\rm x},t)+b$,where $a=0.99925$
and $b=4.41542\times 10^{-4}$;(b)Fit the line $\widehat{f}_i^{(\rm
neq)}({\rm x},t)=af_i^{(\rm neq)}({\rm x},t)+b$,where $a=0.83655$
and $b=-1.51056\times 10^{-8}$. Standard
deviation:(a)$\sigma=3.52548\times 10^{-4}$;(b)$\sigma=1.01264\times
10^{-4}$.} \label{fig.3}
\end{figure}

\begin{figure}
\center{\includegraphics{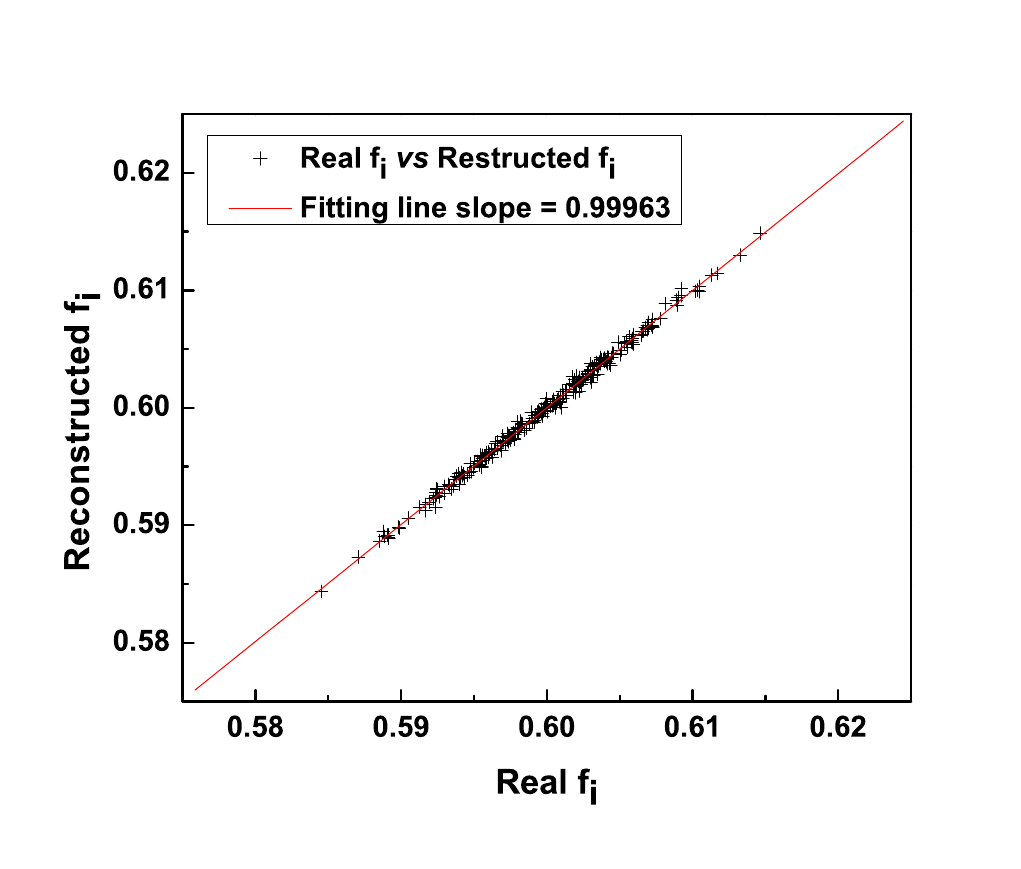}}\center{(a)Linear
regression between $f_i({\rm x},t)$ and $\widehat{f}_i({\rm x},t)$ }
\center{\includegraphics{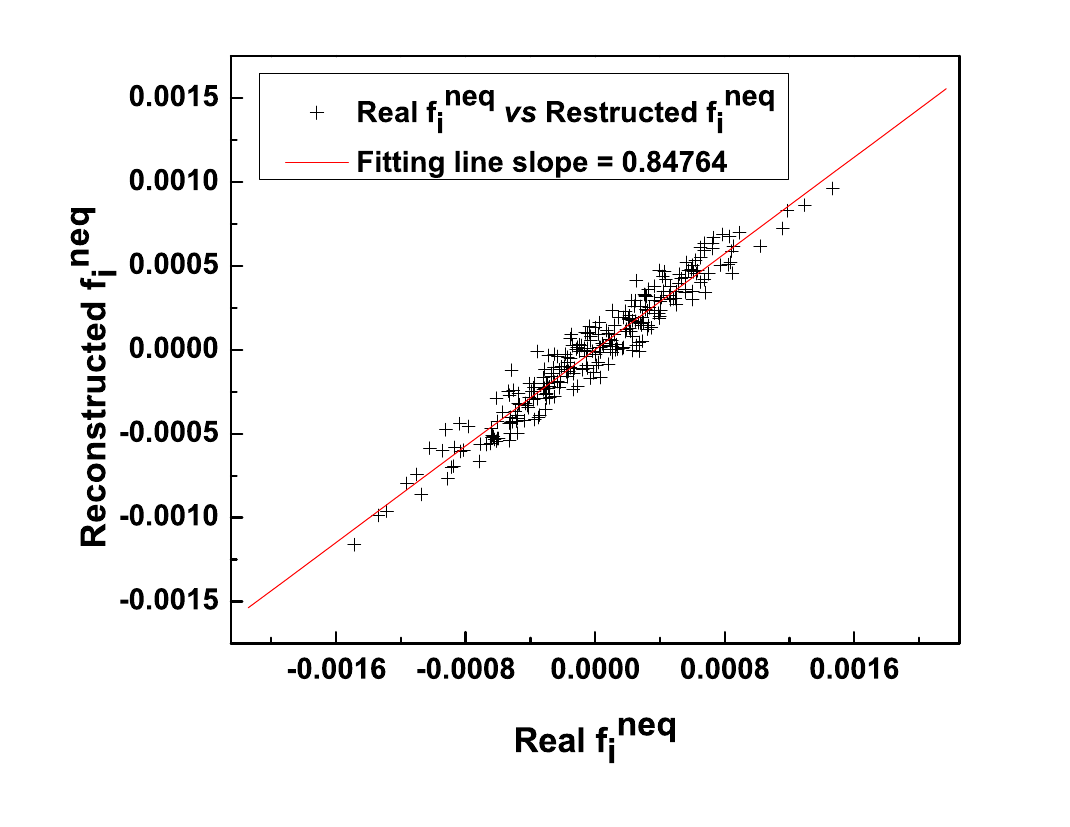}}\center{(b)Linear
regression between $f_i^{(\rm neq)}({\rm x},t)$ and
$\widehat{f}_i^{(\rm neq)}({\rm x},t)$}\caption{Linear regression
(D2Q17, $u_0=0.01$, $t=10000\delta t$, $i=2$): (a)Fit the line
$\widehat{f}_i({\rm x},t)=af_i({\rm x},t)+b$,where $a=0.99963$ and
$b=2.17\times 10^{-4}$;(b)Fit the line $\widehat{f}_i^{(\rm
neq)}({\rm x},t)=af_i^{(\rm neq)}({\rm x},t)+b$,where $a=0.84764$
and $b=-2.17758\times 10^{-8}$. Standard
deviation:(a)$\sigma=3.37821\times 10^{-4}$;(b)$\sigma=9.47431\times
10^{-5}$.} \label{fig.4}
\end{figure}

\begin{figure}
\center{\scalebox{0.7}[0.7]{\includegraphics{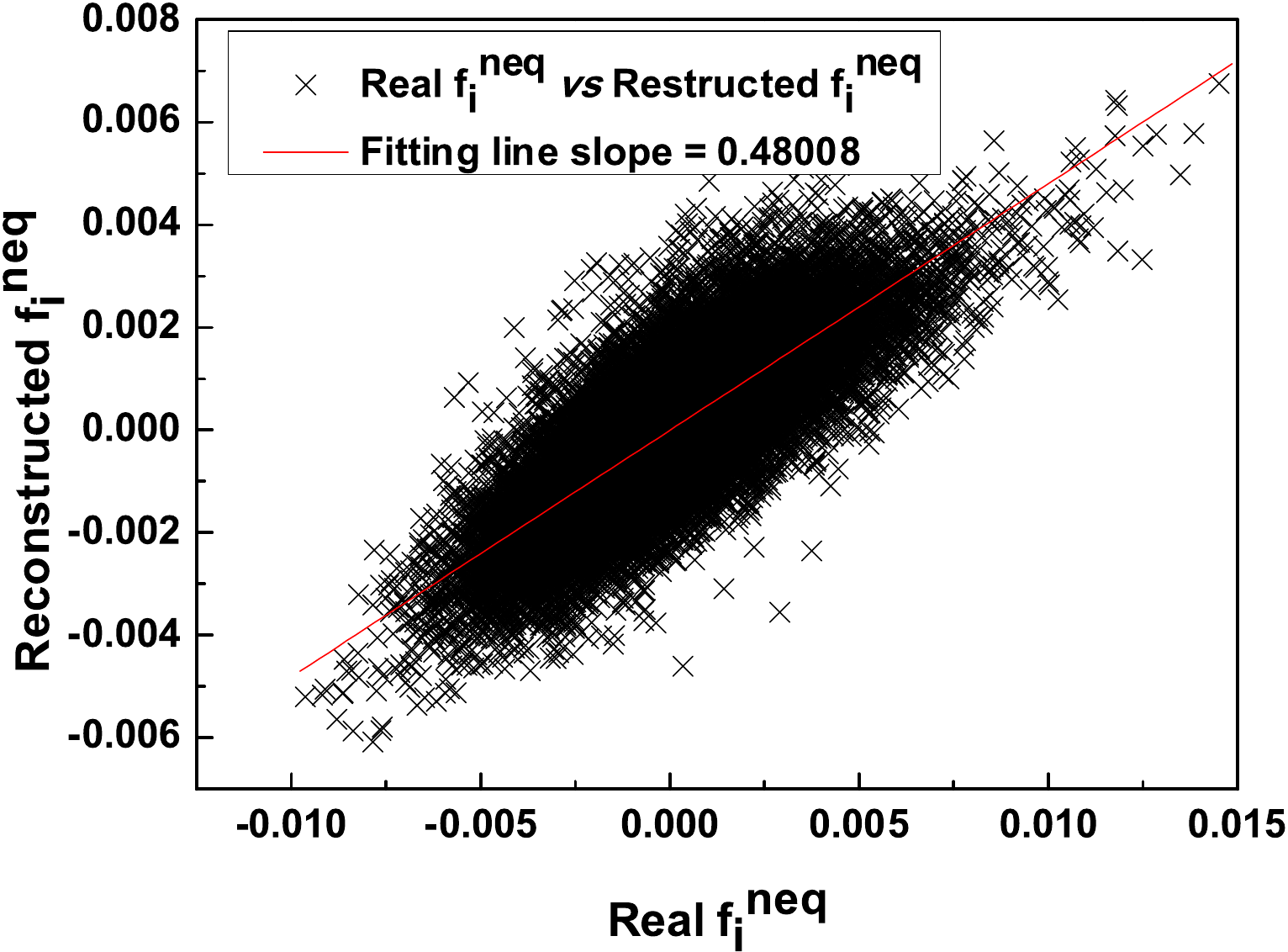}}}
\caption{Linear regression (D2Q9, $u_0=0.1$, $t=1000\delta t$,
$i=2$): Fit the line $\widehat{f}_i({\rm x},t)=af_i({\rm
x},t)+b$,where $a=0.48088$ and $b=-0.248003\times 10^{-6}$.}
\label{fig.guo}
\end{figure}

In Figs (\ref{fig.1})-(\ref{fig.4}), the reconstructed
single-particle distribution functions and non-equilibrium
distribution functions are compared with the real single-particle
distribution functions and non-equilibrium distribution functions by
linear regression analysis. When $u_0=0.1$ and $t=1000\delta t$, it
is clear that the reconstructed single-particle distribution
functions and the non-equilibrium distribution functions coincide
with the real single-particle distribution functions and
non-equilibrium distribution functions very well for D2Q9 and D2Q17
in Figs (\ref{fig.1})-(\ref{fig.2}). The corresponding relative
errors ${{\rm E}(f_i,\widehat{f}_i)}$ are about
$0.242\%$ and $0.194\%$, respectively. The relative errors
${{\rm E}(f_i^{\rm(neq)},\widehat{f}_i^{\rm(neq)})}$ are
about $16.735\%$ and $15.782\%$ for the single-particle
non-equilibrium distribution functions of D2Q9 and D2Q17,
respectively. If Eq.(\ref{Imamuraf}) by Imamura {\it et al}
\cite{Imaura} is used to calculate the single-particle
non-equilibrium distribution functions, the relative errors
${{\rm E}(f_i^{\rm(neq)},\widehat{f}_i^{\rm(neq)})}$ are
up to about $21.65\%$ and $18.13\%$ for D2Q9 and D2Q17,
respectively. We also adopted Eqs. (\ref{skodorsneq}) in
\cite{Skordos} and (\ref{Guo})in \cite{guo} to do the same
calculations. The relative errors ${{\rm
E}(f_i^{\rm(neq)},\widehat{f}_i^{\rm(neq)})}$ of the single-particle
non-equilibrium distribution functions can be up to about $80\%$ at
many lattice nodes. In Fig. \ref{fig.guo}, the
numerical relation between $f_i^{\rm(neq)}$ and
$\widehat{f}_i^{\rm(neq)}$ for the method in \cite{guo}. The mean
relative error ${\rm
E}(f_i^{\rm(neq)},\widehat{f}_i^{\rm(neq)})$ is larger than
$43.74\%$ for D2Q9.  In the statistical procedure, we ignore the
points with very small $f_i^{\rm(neq)}$ and
$\widehat{f}_i^{\rm(neq)}$ ($f_i^{\rm(neq)},
\widehat{f}_i^{\rm(neq)}<10^{-3}$) for the method in \cite{guo}.
Here, we must point out that when $f_i^{\rm(neq)}$ and
$\widehat{f}_i^{\rm(neq)}$
 are very small, the relative errors ${\rm E}(f_i^{\rm(neq)},\widehat{f}_i^{\rm(neq)})$ of the methods in \cite{Skordos,guo} are very large.
 In such a circumstance, the relative error of the
non-equilibrium distribution functions by Eq.
(\ref{lbmthermalvelocity}) is also a bit larger, but it still less
than that computed by Eq. (\ref{Imamuraf}) \cite{Imaura} and much
less than that computed by Eqs. (\ref{skodorsneq})$\sim$ (\ref{Guo})
of \cite{Skordos} and \cite{guo}, respectively. Similar results can
be observed for the case of $u_0=0.01$ at $t=10000\delta t$ for D2Q9
and D2Q17. For the simplicity of presentation, they are not provided
here.

 In addition we also found that when the single-particle
distribution functions and non-equilibrium distribution functions
are reconstructed, the results from D2Q17 model show a better
accuracy than that of D2Q9 model. Meanwhile, from the both models,
more accurate results can be gained when the Mach number is reduced.
Such results are very reasonable, and can be understood as follows.
First, D2Q17 model is more accurate to approach Maxwell distribution
function in discrete velocity spaces than D2Q9 model. Second, low
Mach number will lead to a reduction of the truncated errors for
approaching Maxwell distributions and a better recovering
Navier-Stokes equation. It is proved \cite{qian2} that D2Q17 model
can eliminate the third-order term of statistical velocity in
recovered Navier-Stokes equation.

\begin{figure}
\centering
\scalebox{0.8}[0.8]{\includegraphics{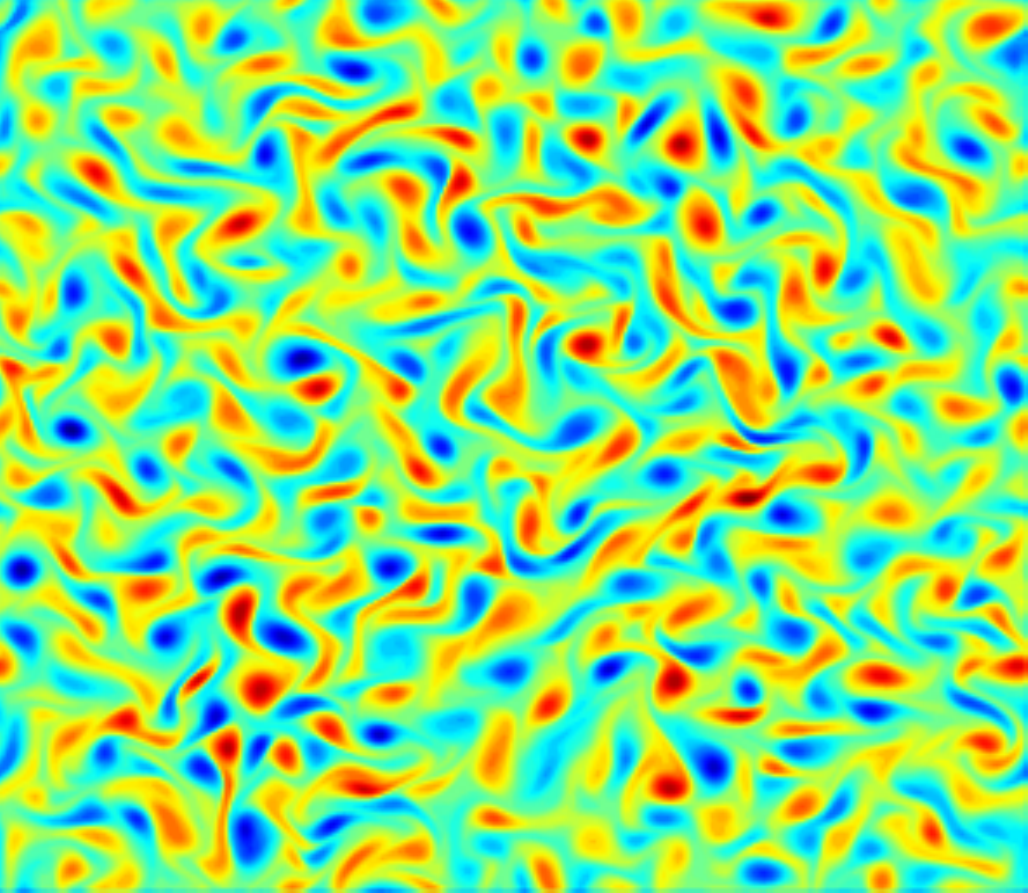}}\center{(a)Vorticity contour plot by the real $f_i({\rm x},t)$ }\\
\scalebox{0.8}[0.8]{\includegraphics{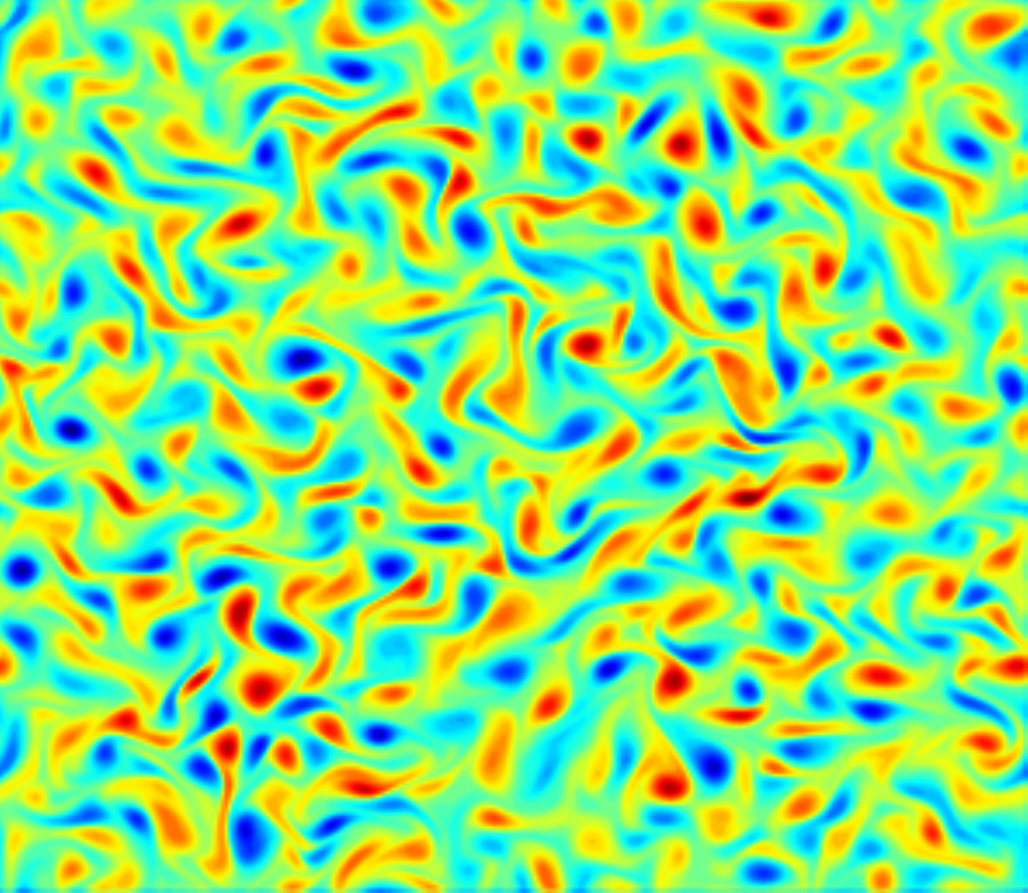}}\center{(b)Vorticity
contour plot by the reconstructed $\widehat{f}_i({\rm
x},t)$}\caption{Vorticity contour plots (D2Q9, $u_0=0.1$,
$t=1000\delta t$): (a)Vorticity contour plot by the real $f_i({\rm
x},t)$ ; (b)Vorticity contour plot by the reconstructed
$\widehat{f}_i({\rm x},t)$} \label{fig.5}
\end{figure}

\begin{figure}
\centering
\scalebox{0.8}[0.8]{\includegraphics{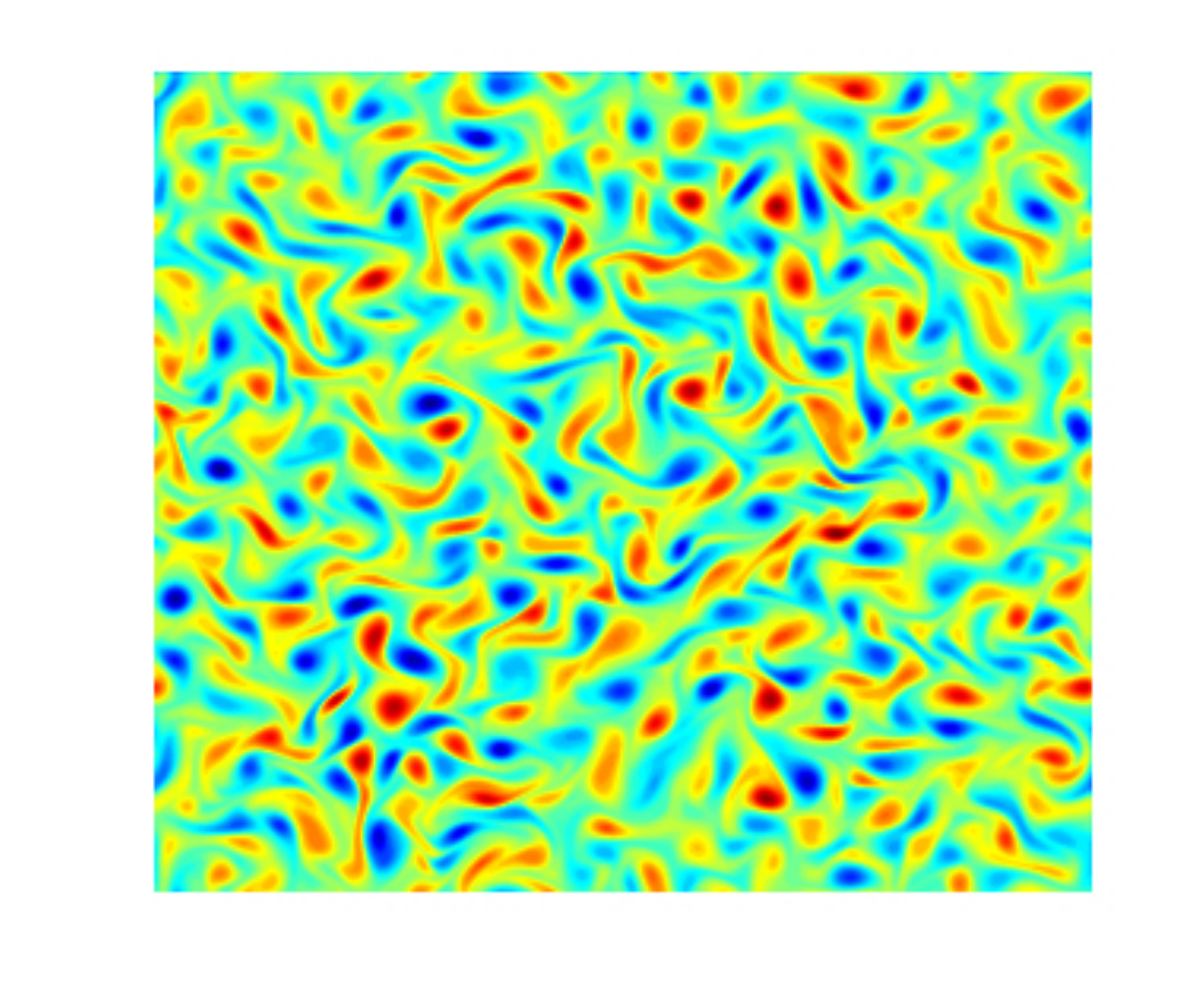}}\center{(a)Vorticity contour plot by the real $f_i({\rm x},t)$ }\\
\scalebox{0.8}[0.8]{\includegraphics{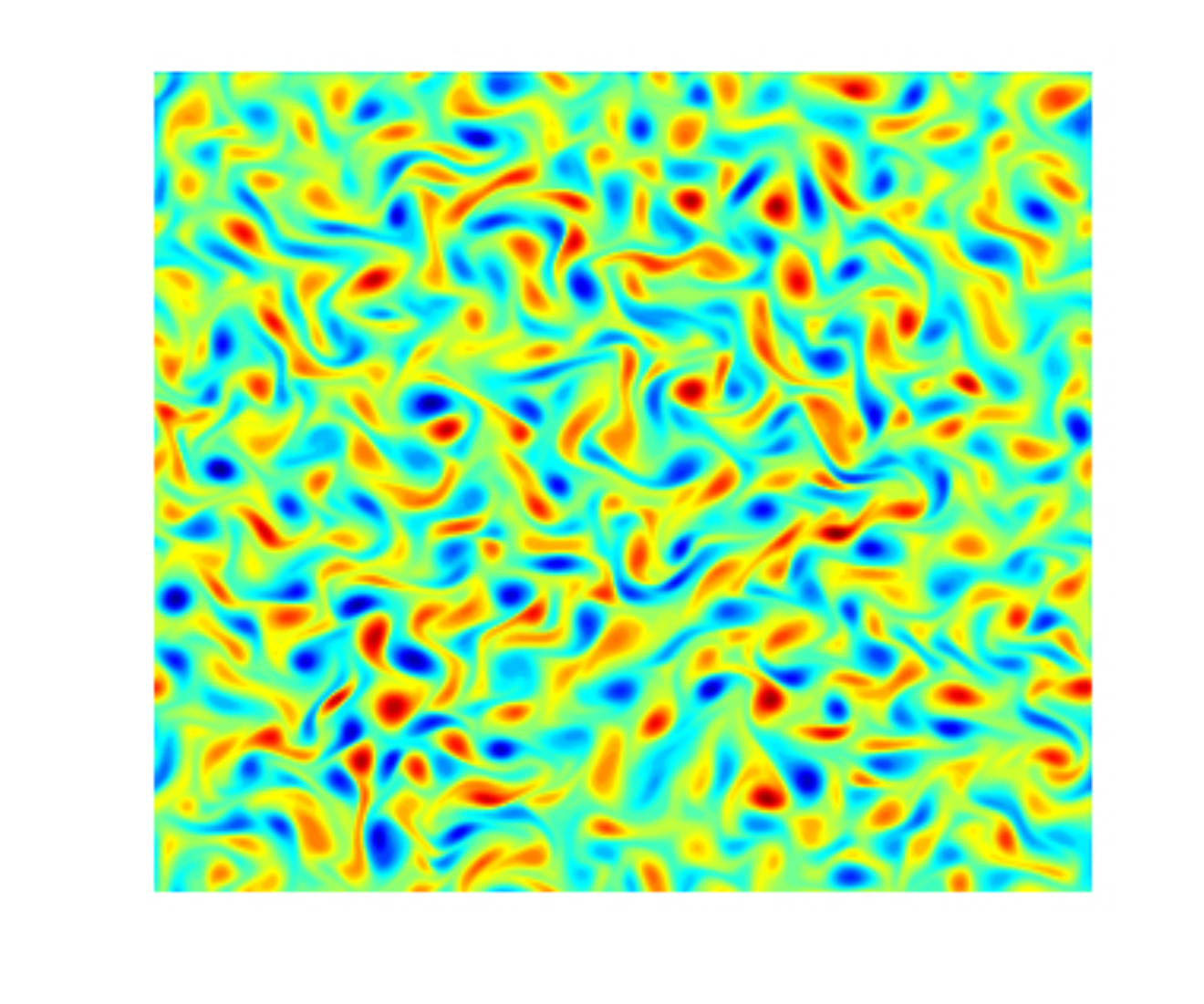}}\center{(b)Vorticity
contour plot by the reconstructed $\widehat{f}_i({\rm
x},t)$}\caption{Vorticity contour plots (D2Q9, $u_0=0.01$,
$t=10000\delta t$): (a)Vorticity contour plot by the real $f_i({\rm
x},t)$ ; (b)Vorticity contour plot by the reconstructed
$\widehat{f}_i({\rm x},t)$} \label{fig.6}
\end{figure}

Finally, attention is turned to the comparison of vorticity by the
real ${f}_i({\rm x},t)$ and the reconstructed $\widehat{f}_i({\rm
x},t)$ in Figs.~\ref{fig.5}$\sim$\ref{fig.6}, where the vorticity
contour figures are given for $u_0=0.1$ and $u_0=0.01$,
respectively.In order to show the quantitative sense
of the vorticity reconstruction error, we choose 100 and 1000
time-series samples for $u_0=0.1$ and $u_0=0.001$, respectively. The
$L^2$-relative departures of the reconstructed vorticity are
$0.02\%\pm 0.0014\%$ (D2Q9, $u_0=0.1$), $0.005\%\pm0.0003\%$ (D2Q17,
$u_0=0.1$), $0.01\%\pm0.0026\%$ (D2Q9, $u_0=0.01$) and $0.003\%\pm
0.0005\%$ (D2Q17, $u_0=0.01$). The agreement is very good.

In all, the proposed two operators can reconstruct the
single-particle distribution functions and non-equilibrium
distribution functions accurately and effectively. It can be shown
that the two reconstruction operators are very flexible to apply to
other discrete velocity models of lattice Boltzmann equation.

\subsection{The rates of convergence }\label{convergence}
\begin{figure}
\centering
\scalebox{0.6}[0.6]{\includegraphics{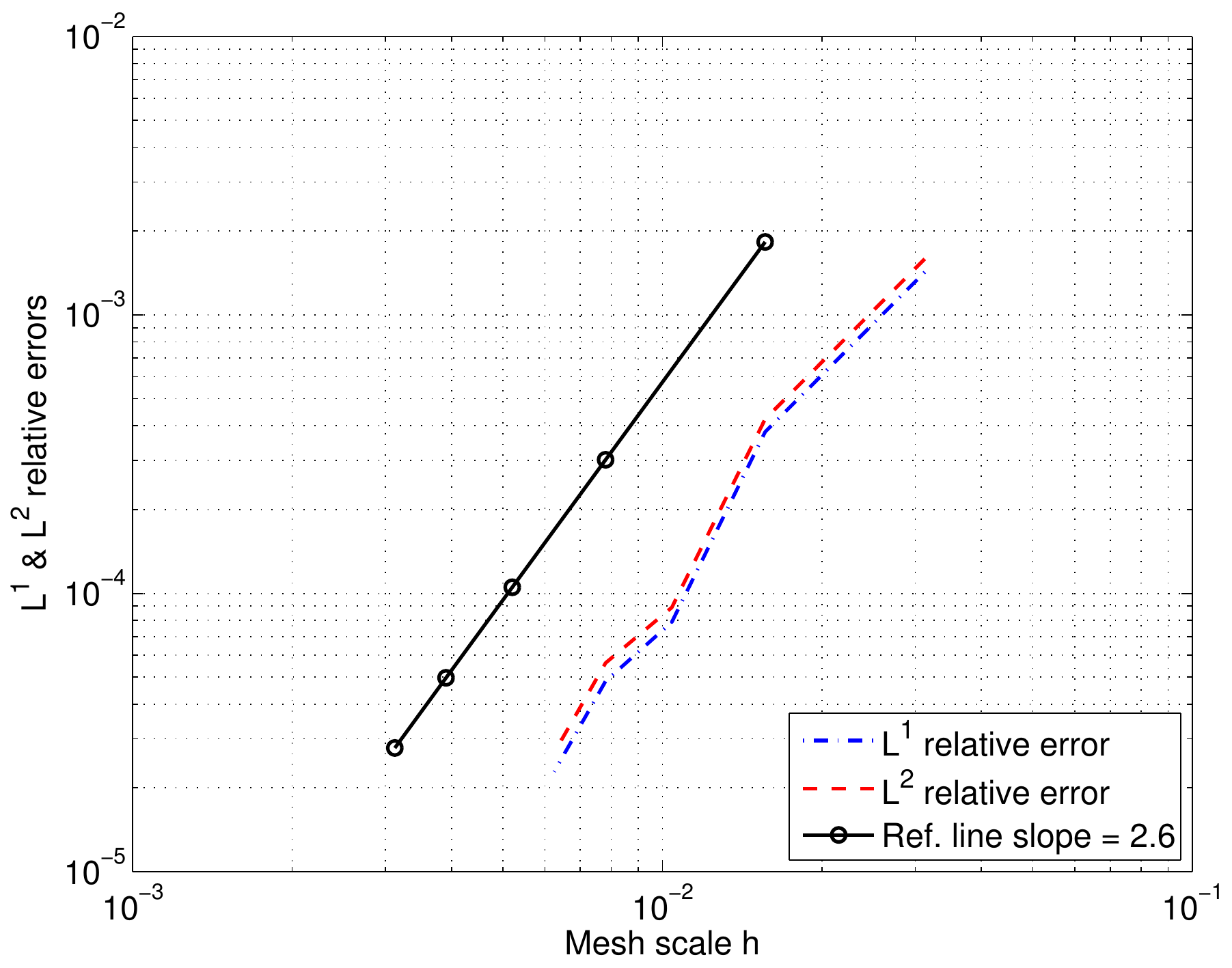}}
\center{(a) the D2Q9 model}\\
\scalebox{0.6}[0.6]{\includegraphics{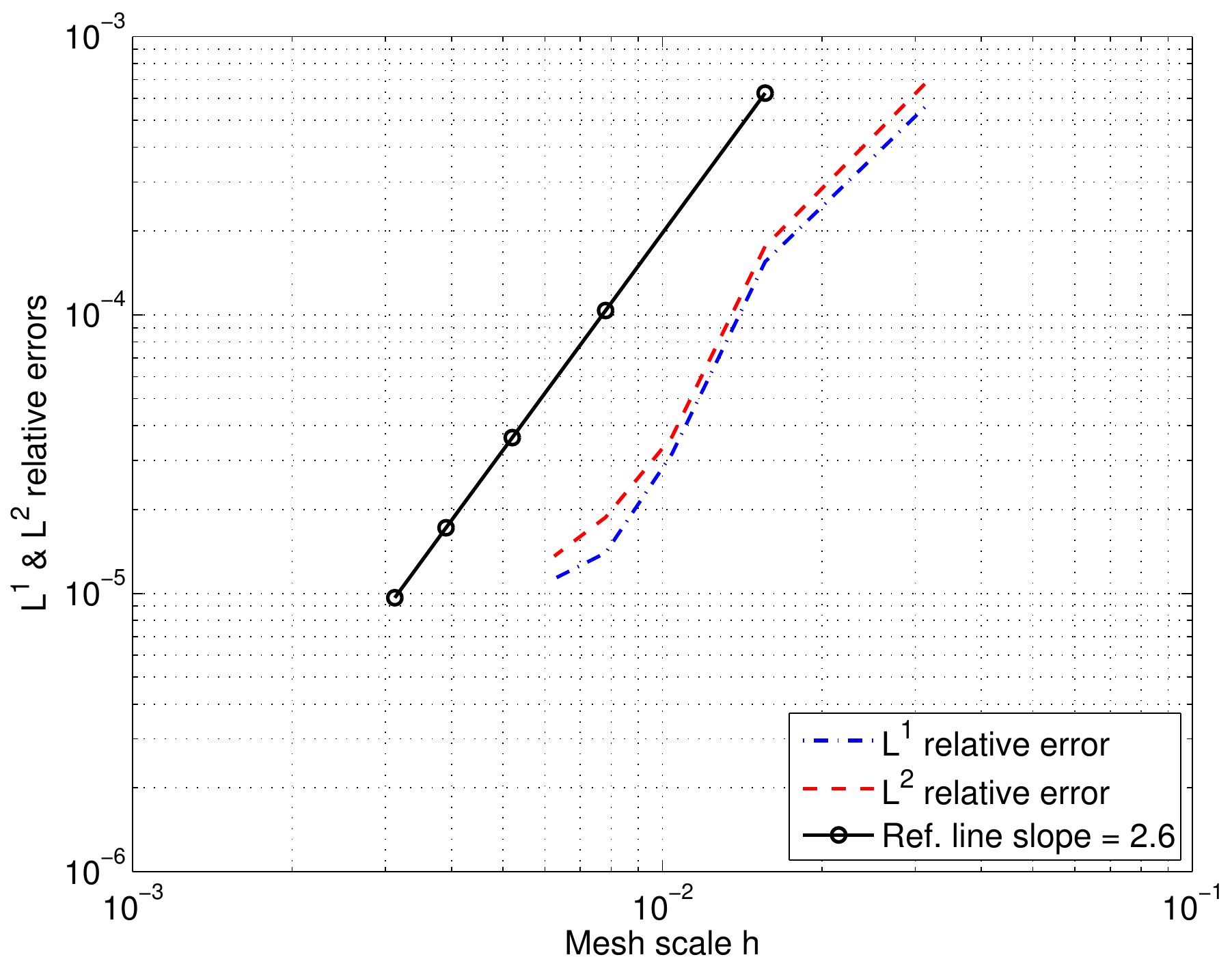}}
\center{(b) the D2Q17 model}\caption{Convergence rates of the reconstruction operator for D2Q9 and D2Q17.} \label{fig.l1l2}
\end{figure}

In order to validate the approach behaviors versus different grid sizes, we give the convergence properties of the D2Q9 and D2Q17 models by different mesh scales.  The 2D Taylor-Green vortex problem is chosen as the intial fields
\begin{equation}
\left\{ \begin{array}{ll}
u=-&A{\rm cos}(k_1x){\rm sin}(k_2y)F(t)\\
v= &A\frac{k_1}{k_2}{\rm sin}(k_1x){\rm cos}(k_2y)F(t)\\
p=&p_0-\dfrac{A^2}{4}\left[{\rm cos}(2k_1x)+\dfrac{k_1^2}{k_2^2}{\rm cos}(2k_2y)\right]F^2(t)
\end{array}\right.
\end{equation}
where $F(t)={\rm exp}\left[-\nu (k_1^2+k_2^2) t\right]$, $A=0.1$, $k_1=k_2=4$ and $p_0=\rho_0c_s^2$. The computational domain $\Omega=[0,2\pi]^2$ and $Re = 10000$.  The periodic boundary conditions are applied in both directions.  The initial distribution functions are initialized by the reconstruction operator. The reconstruction $L^1$ and $L^2$ relative errors of the distribution functions are calculated at the time steps $n=\{2000, 4000, 6000, 8000, 10000\}$ corresponding to the mesh resolutions $h=\{1/32,1/64,1/96,1/128,1/160\}$ respectively.  In Fig. \ref{fig.l1l2}, the relative errors are given in the log-log coordinates. From the results, it is clear that for the D2Q9 model and the D2Q17 model, they nearly have the same convergence rates which are approximately equal to 2.6. However, the relative errors of the D2Q17 model are smaller than that of the D2Q9 model.  That means the reconstruction precision can be improved when the number of the discrete velocity increases. This conclusion is consistent with the result in Sec. \ref{turb}.

\subsection{Coupling computations of FVM and LBM for lid-driven cavity flows}
\begin{figure}
\centering
\scalebox{0.4}[0.4]{\includegraphics{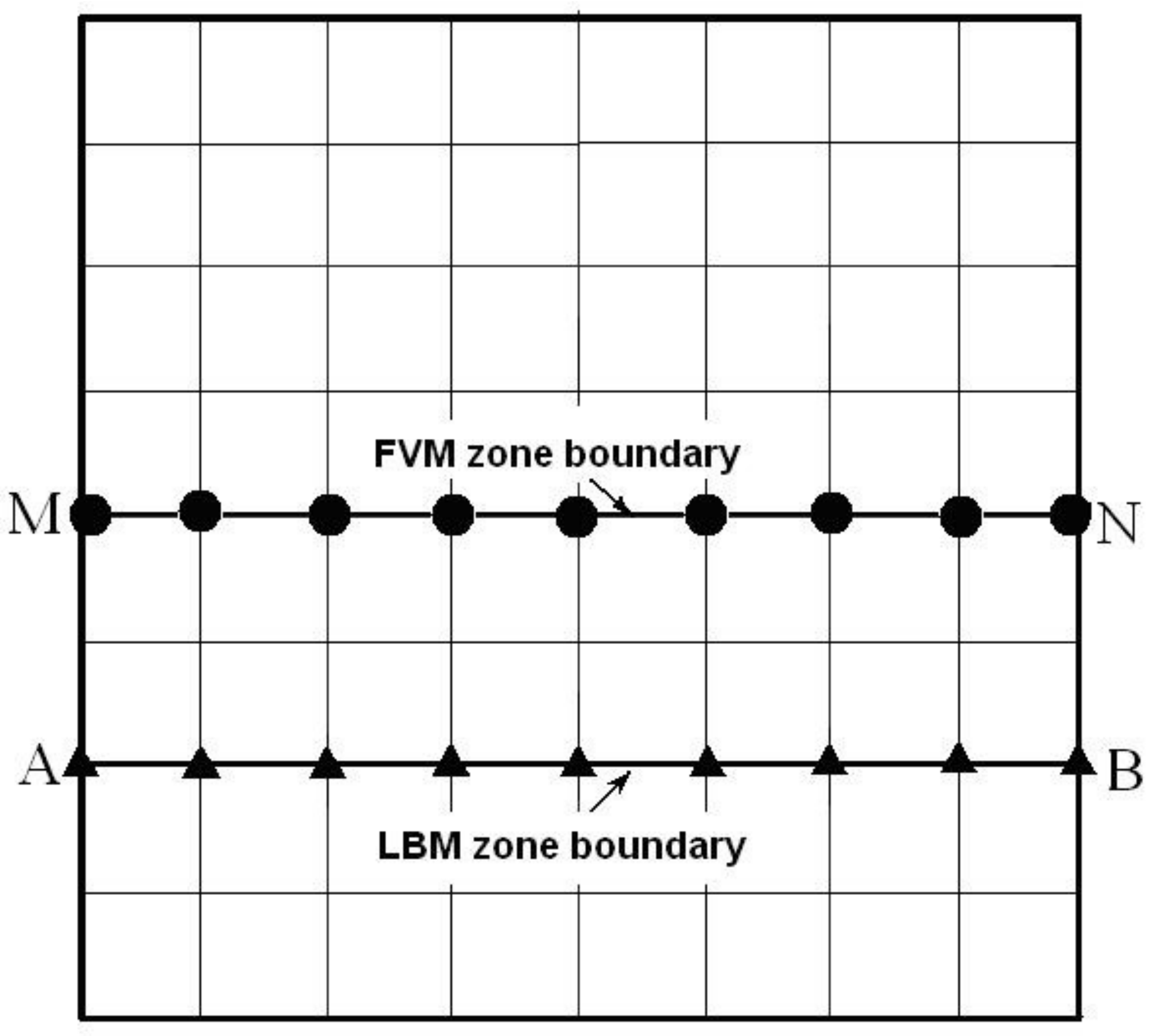}}
\center{(a)Interface structure between two regions of FVM and LBM}\\
\scalebox{0.4}[0.4]{\includegraphics{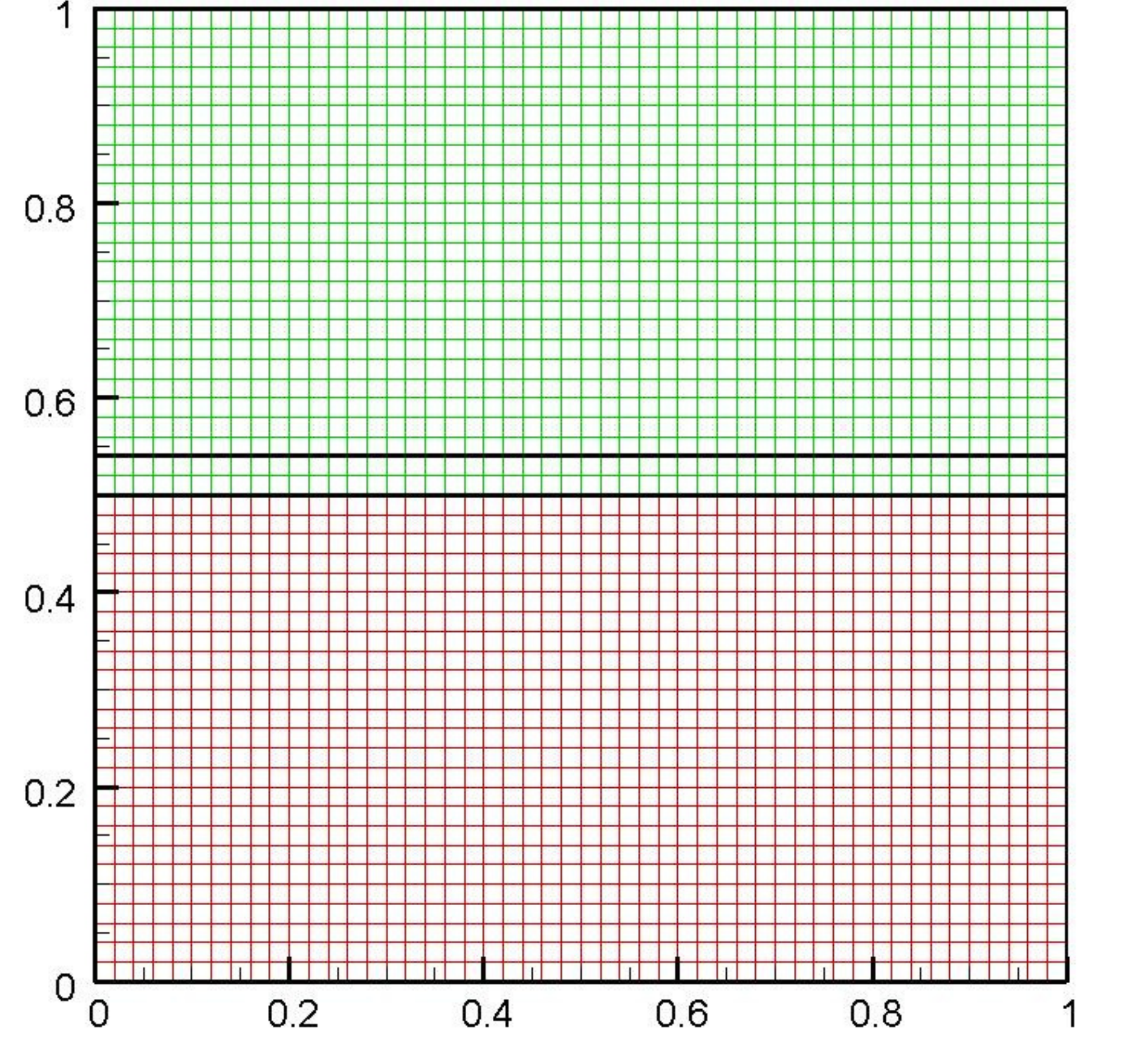}}
\center{(b) Grid layout for a 2D lid-driven cavity ($200\times
200$)}\caption{Geometric structure and mesh partition: (a)Interface
structure between two regions of FVM and LBM; (b)Grid layout for a
2D lid-driven cavity ($200\times 200$)} \label{fig.7}
\end{figure}

In order to illustrate the feasibility of the recommended
reconstruction operator, the lid-driven cavity flow is simulated by
the coupled LBM-FVM method. The computational domain is decomposed
in two regions in which the LBM and FVM methods are used
respectively (see Fig.~\ref{fig.7}-(a)). The coarseness and fineness
of the grids can adjusted according to the zone spatial scale in
each region. If the grid systems at the interface of overlap
subregions are not identical, space interpolation at the interface
is required when transferring the information at the interface. In
this paper, the identical mesh structures are used for FVM and LBM
for convenience to avoid the spatial interpolation (see
Fig.~\ref{fig.7}-(b)). In order to implement the coupling
computations, the overlap Schwartz alternative procedure is used to
handle the computations.

\begin{figure}
\centering
\scalebox{0.2}[0.2]{\includegraphics{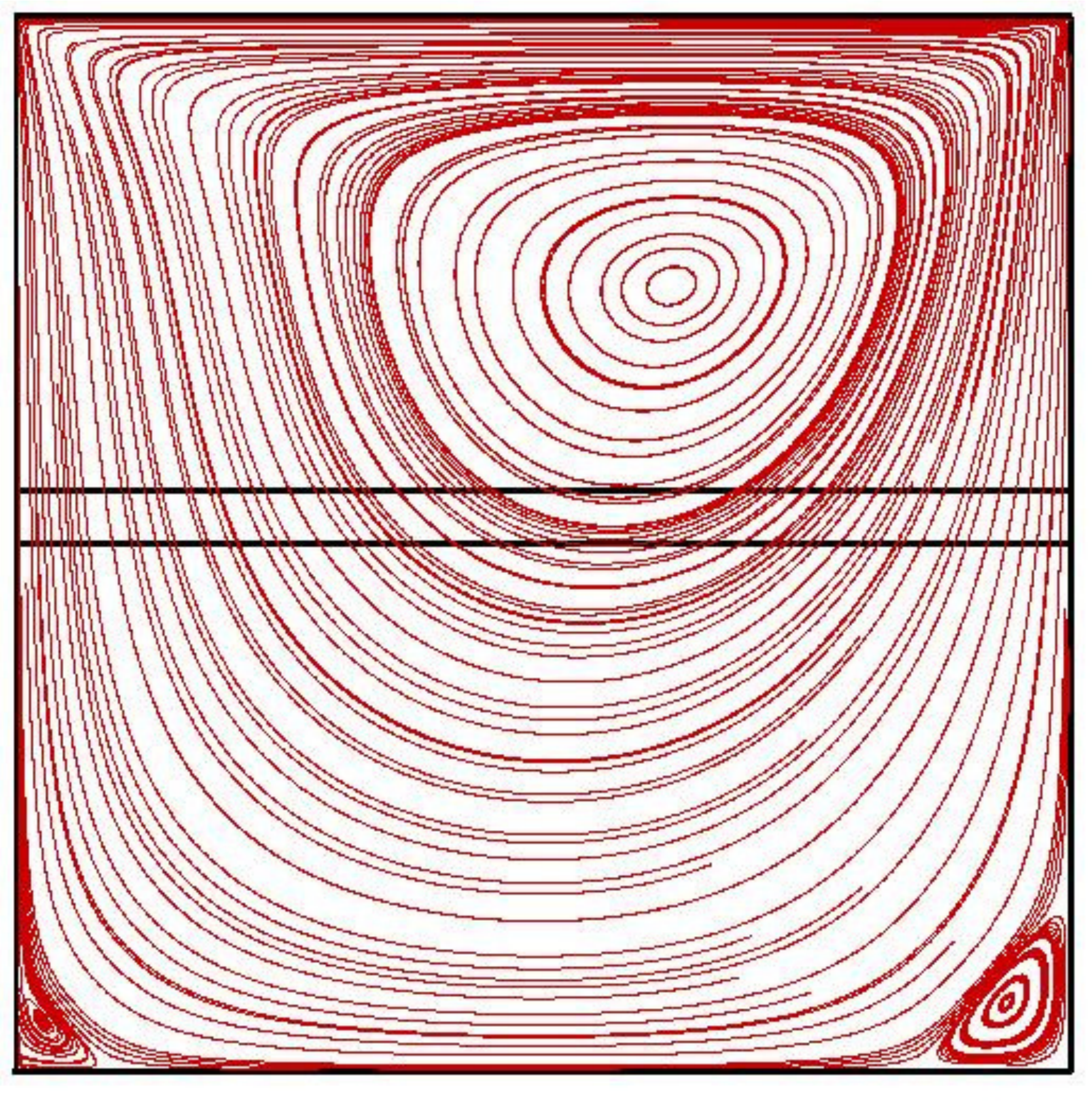}}
\scalebox{0.2}[0.2]{\includegraphics{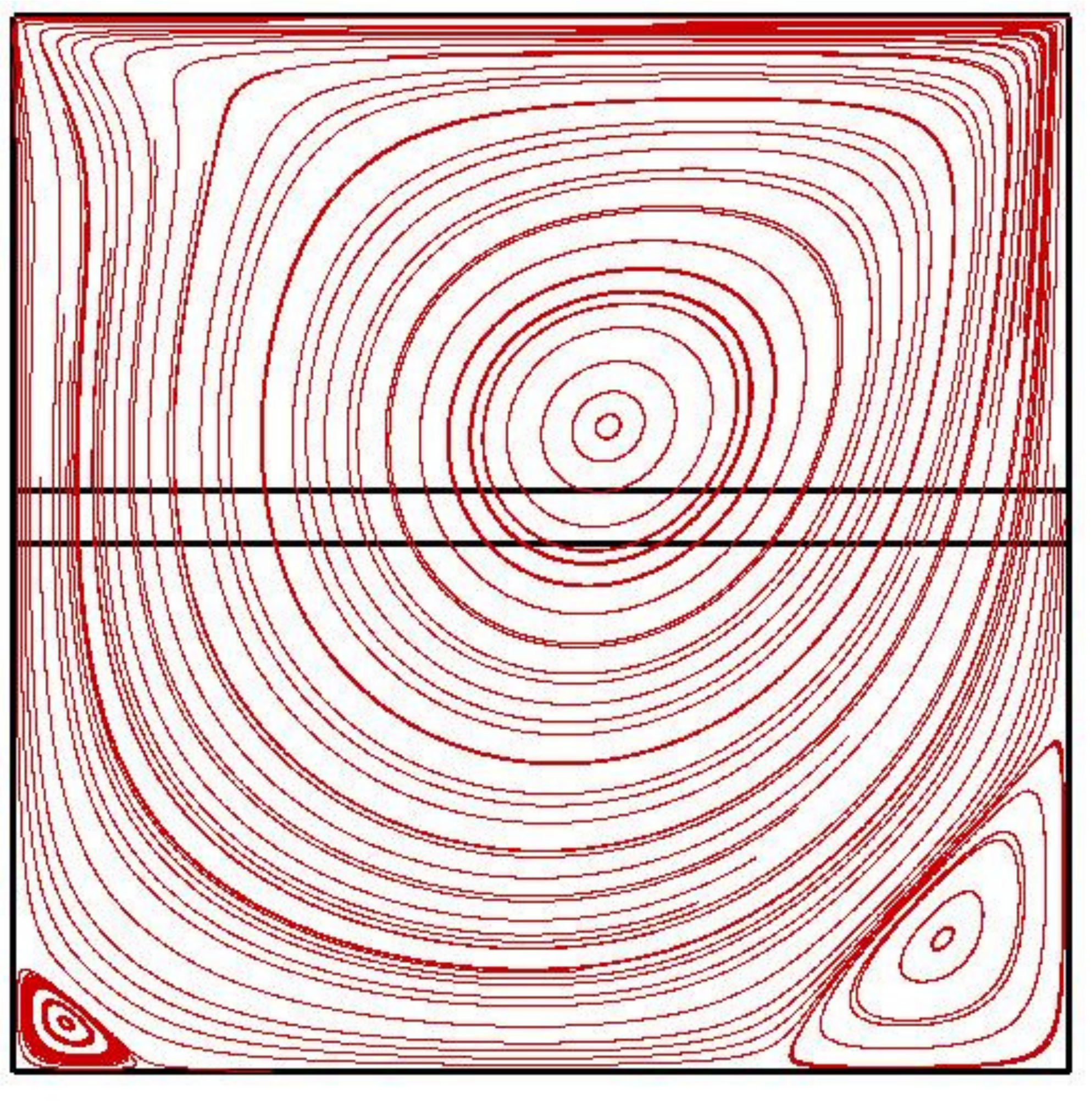}}
\scalebox{0.2}[0.2]{\includegraphics{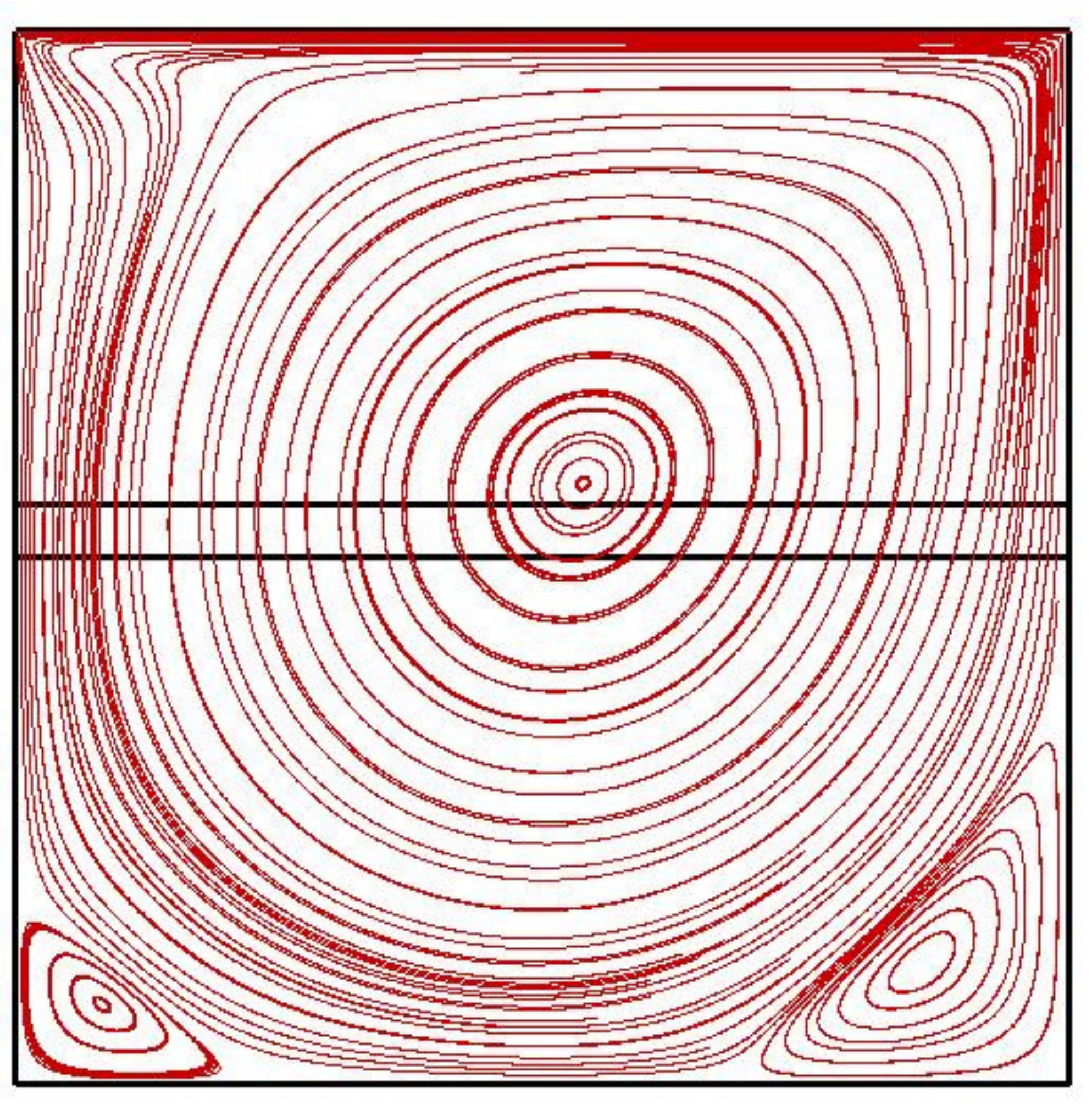}}\center{(a)
$Re=100$ \qquad\qquad(b) $Re=400$ \qquad\qquad(c) $Re=1000$}
\caption{ Contour plots of streamline for different Reynolds
numbers} \label{fig.8}
\end{figure}

Numerical simulations were carried out for cavity flow of $Re=100,
400$ and $1000$ on a grid $200\times200$. The characteristic length
of square cavity is $L=1$. The boundaries of the cavity are
stationary walls, except the top-boundary with a uniform tangential
velocity ($u_{t,Re=100}=3.33\times 10^{-3}$,
$u_{t,Re=400}=1.33\times 10^{-3}$, $u_{t,Re=1000}=3.33\times
10^{-2}$). Fig.~\ref{fig.8} shows plots of the stream function for
the Reynolds number considered. These plots give a clear picture of
the overall flow pattern and the effect of Reynolds number on the
structure of the recirculating eddies in the cavity. The smoothness
of the stream function distribution, especially around the overlap
region confirms the correctness of the information transfer at the
interface. To further quantify these results, the velocity profiles
along the vertical and horizontal centerlines of the cavity are
shown in Fig.~\ref{fig.9}. The results are in close agreement with
the benchmark solution \cite{Ghia}. The smoothness and consistency
of velocity distribution in the overlap region is presented in
Fig.~\ref{fig.10} where a local, enlarged view of the vector plot in
the overlap region is shown. Clearly, the vectors in the overlap
region are quiet consistent between the LBM results and the FVM
results. Figs.~\ref{fig.11} and \ref{fig.12} show the contours of
horizontal and vertical velocity. It is seen that these physical
quantities are all smooth across the interface. According to the
authors' numerical experience, the smoothness of
vorticity contour is the most difficult to obtain for
such coupled computation, because vorticity if the derivative of
velocity. The contours of vorticity distribution are shown in
Fig.~\ref{fig.13}. Over all, the smoothness on the overlap region
are quite good, with a minor bumpiness of the left-hand vortex
contours for the case of $Re=100$.

In all, by the proposed lifting relation, we can couple the
mesoscopic LBM with FVM to implement the domain decomposition
coupling-computations. This paves the way for implementing
multiscale computations based on LBM and macro-numerical methods of
finite-family.

It should be noted that we also tried the coupling computations
based on the distribution function $f_i({\rm x},t)$ reconstructed by
Eq. (\ref{skodorsneq}) of \cite{Skordos} and (\ref{Guo}) of
\cite{guo}. Unfortunately, all of our tries were unsuccessful and
converged solutions could not be obtained.

\section*{Conclusion}
In this paper, we derive the relation to lift the macroscopic
variables to the microscopic variables for LBM. Two methods of
derivation are conducted and they lead to the same result. Numerical
tests demonstrate that the derived lifting relation possesses good
precision. The proposed lifting relation offers a way to implement
the multiscale-computations involving LBM more efficiently and
robustly.

\section*{Acknowledgment}
 This work was supported by the Key Projects National Natural Science Foundation of China (51136004)
 and the National Basic Research Program (973) (2007CB206902). We appreciate the referee's valuable comments on our work.

\newpage \clearpage
\newpage\clearpage

\begin{figure}
\centering\scalebox{0.5}[0.5]{\includegraphics{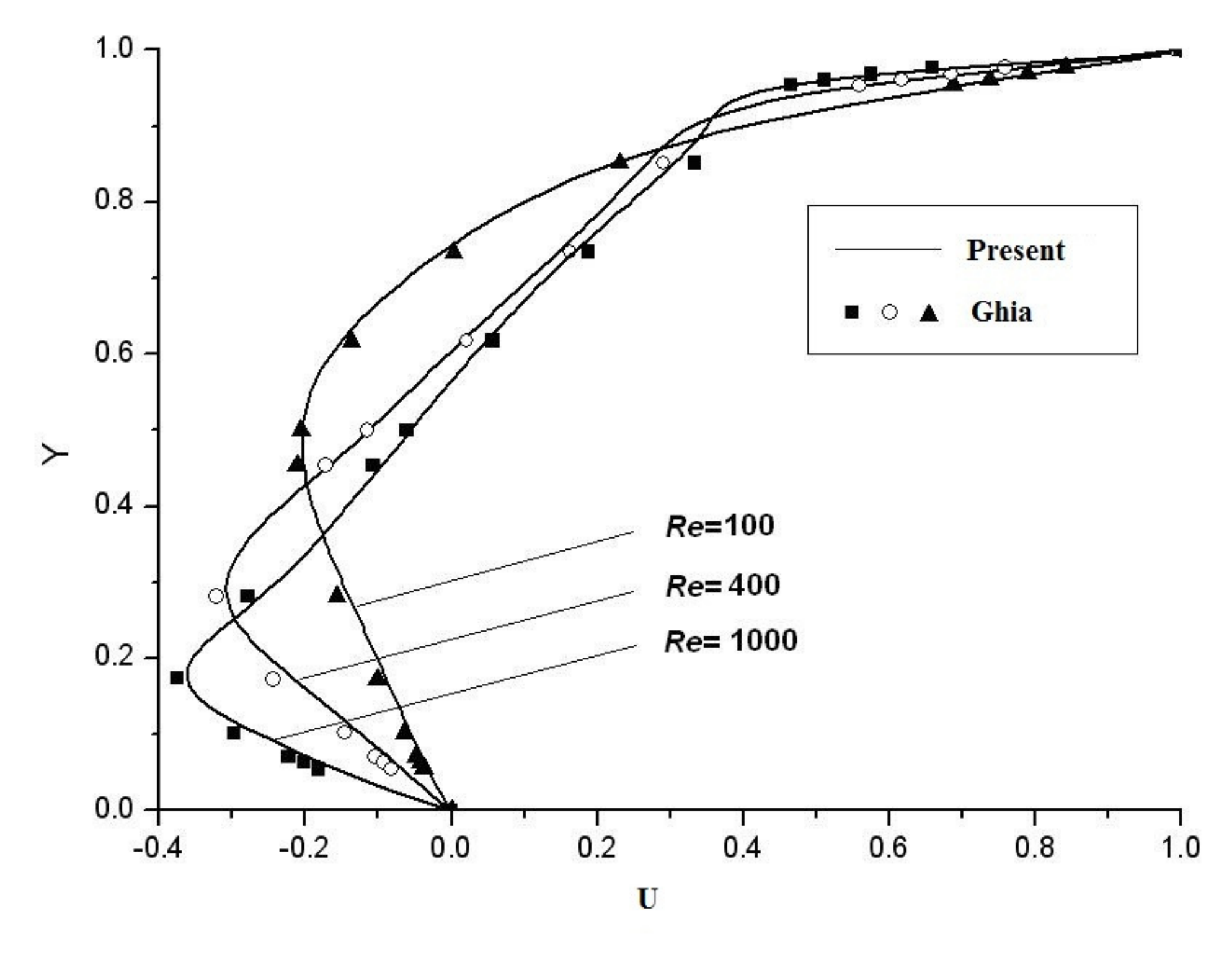}}
\center{(a) Horizontal velocity profiles }
\end{figure}
\begin{figure}
\centering\scalebox{0.5}[0.5]{\includegraphics{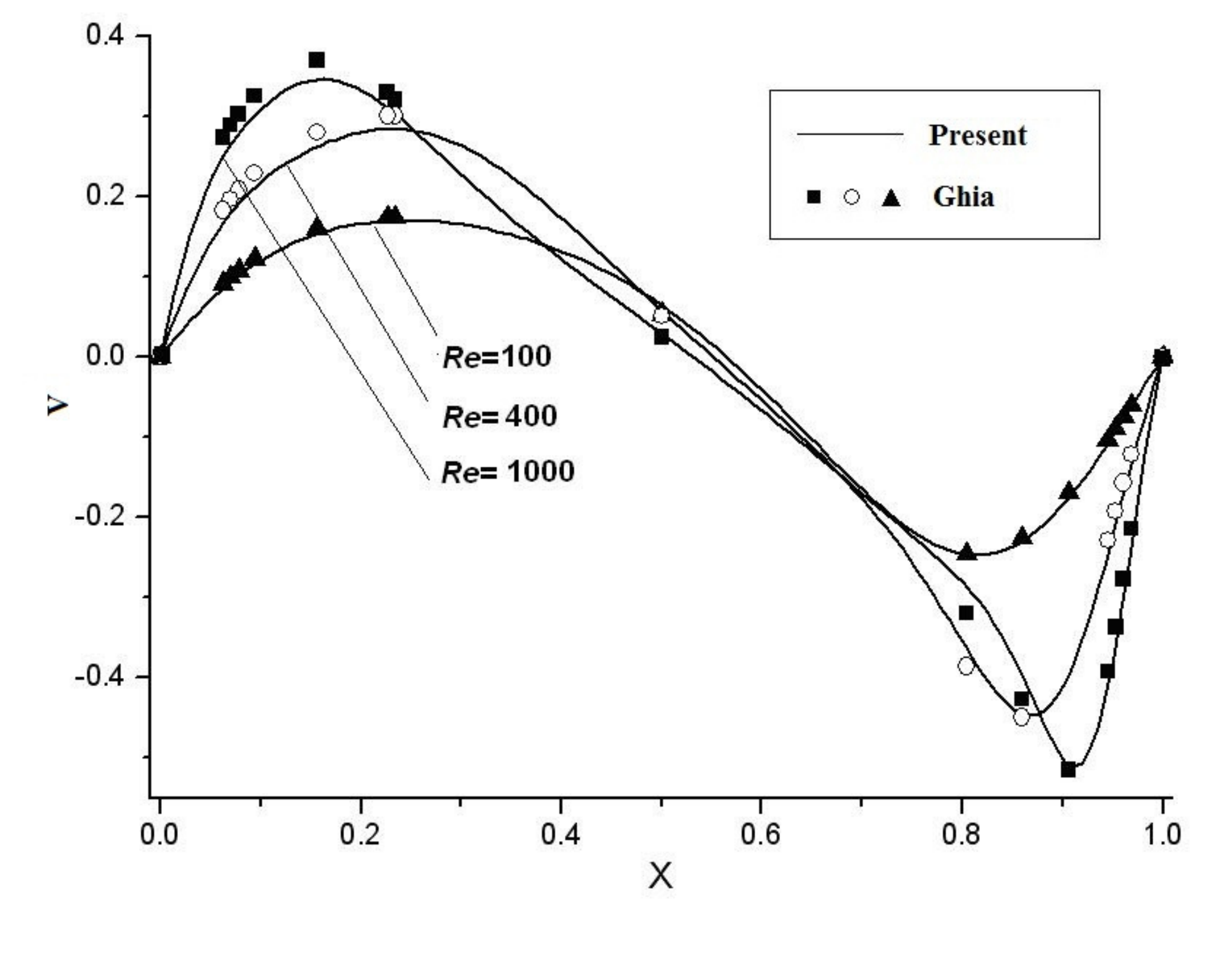}}
\center{(b)Vertical velocity profiles }\caption{ Comparisons between
Ghia's benchmark solutions and coupling solutions} \label{fig.9}
\end{figure}

\begin{figure}
\centering\scalebox{0.7}[0.7]{\includegraphics{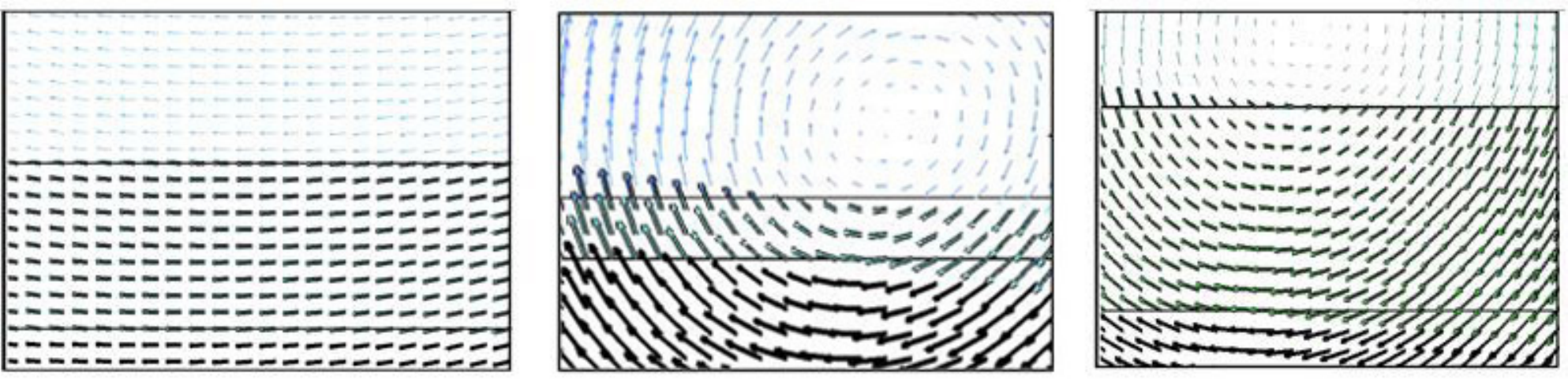}}
\center{(a) $Re=100$ \qquad\qquad\qquad(b) $Re=400$
\qquad\qquad\qquad(c) $Re=1000$}\caption{Enlarge vector plots in
overlap regions} \label{fig.10}
\end{figure}

\begin{figure}
\centering
\scalebox{0.2}[0.2]{\includegraphics{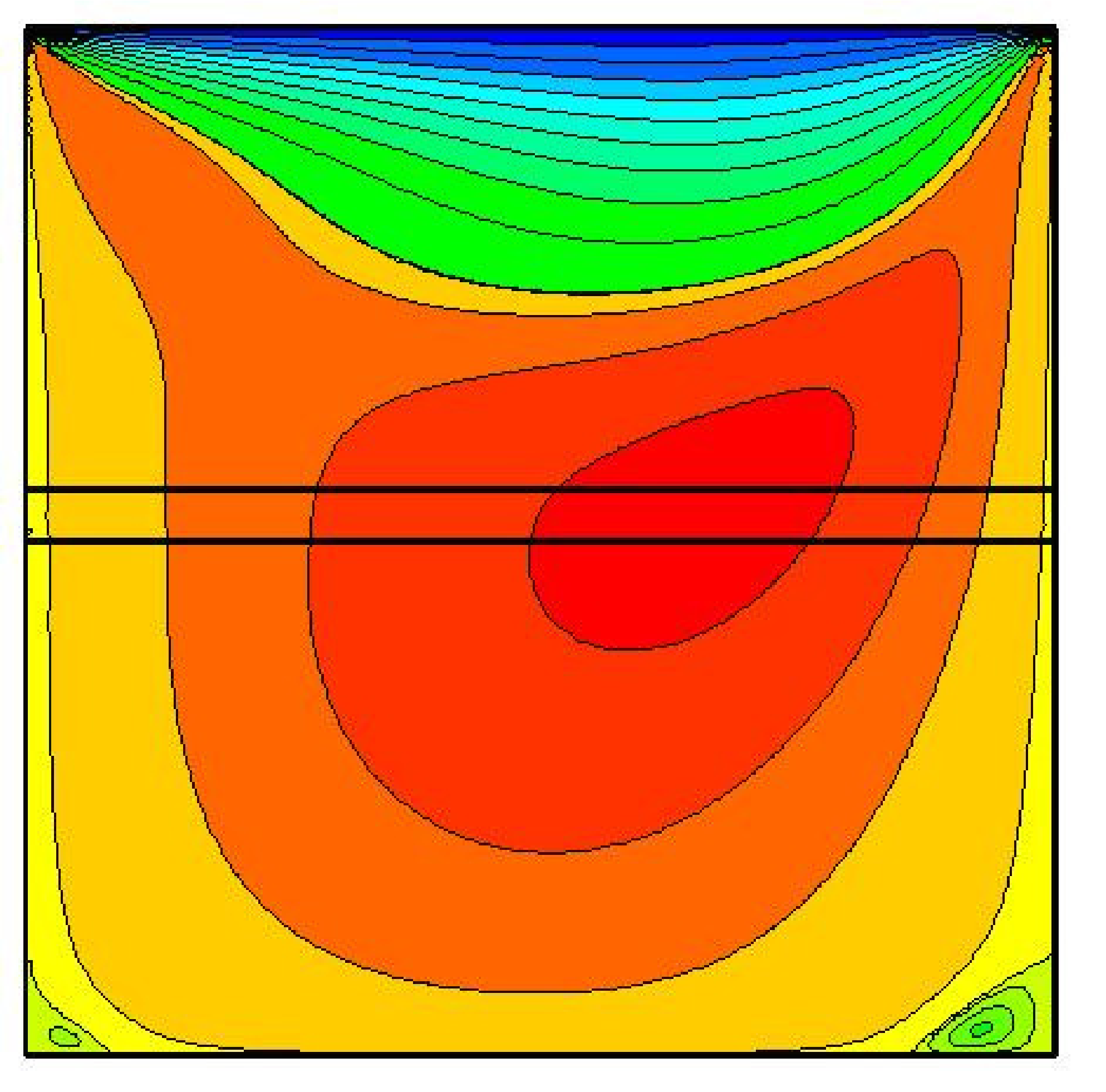}}
\scalebox{0.2}[0.2]{\includegraphics{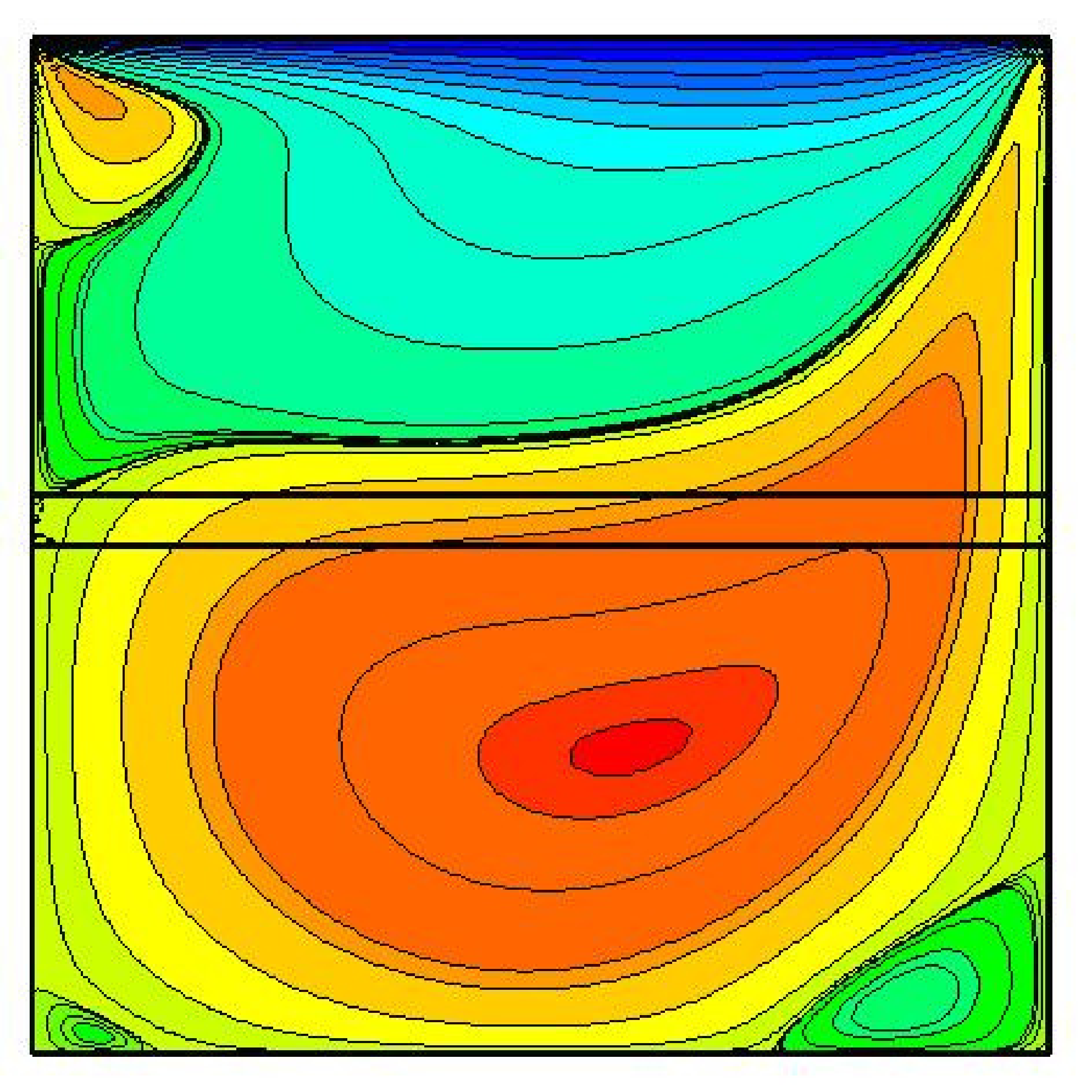}}
\scalebox{0.2}[0.2]{\includegraphics{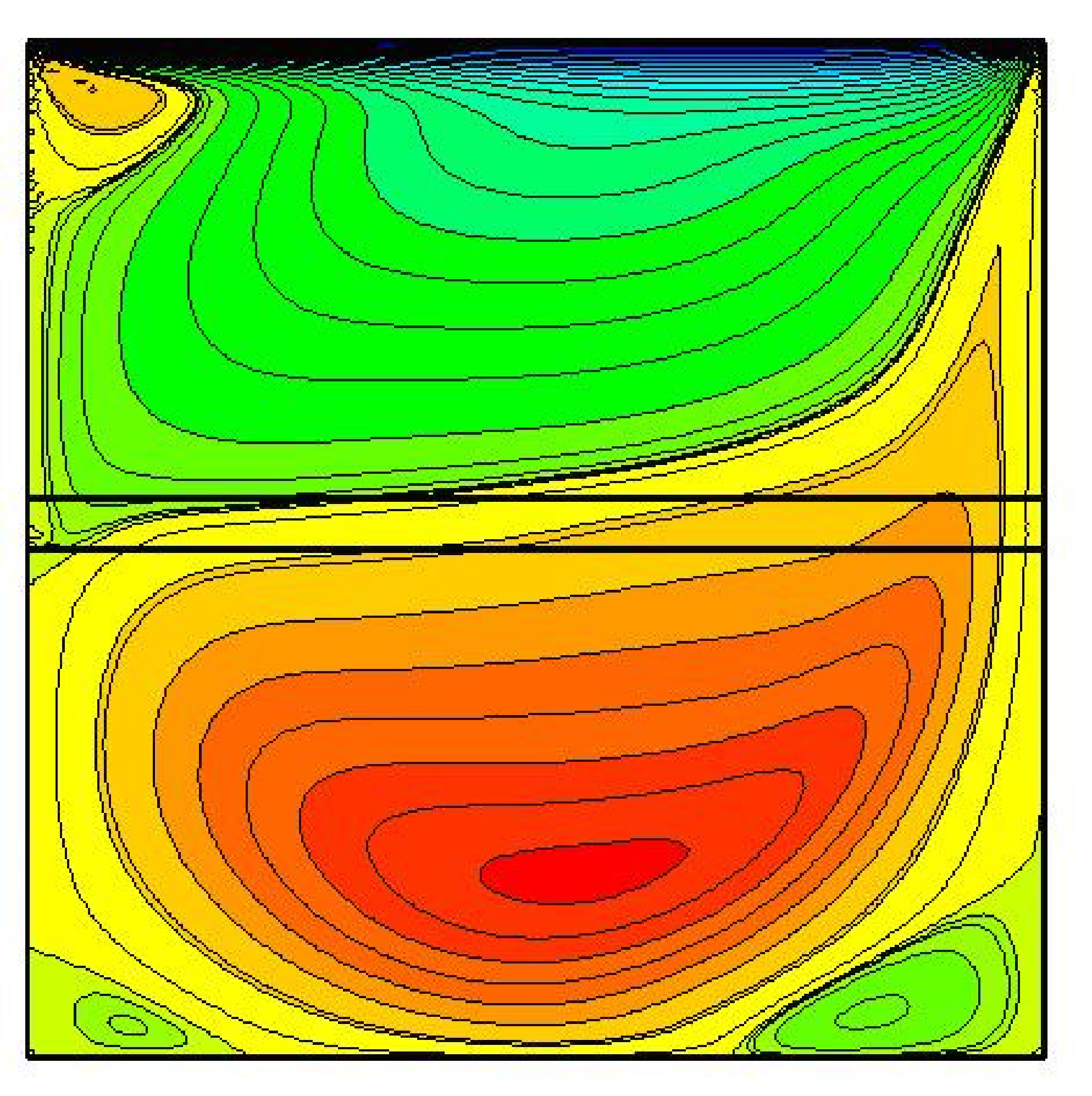}}\center{(a)
$Re=100$ \qquad\qquad(b) $Re=400$ \qquad\qquad(c) $Re=1000$}
\caption{ Contour plots of horizontal velocity for different
Reynolds numbers} \label{fig.11}
\end{figure}

\begin{figure}
\centering
\scalebox{0.2}[0.2]{\includegraphics{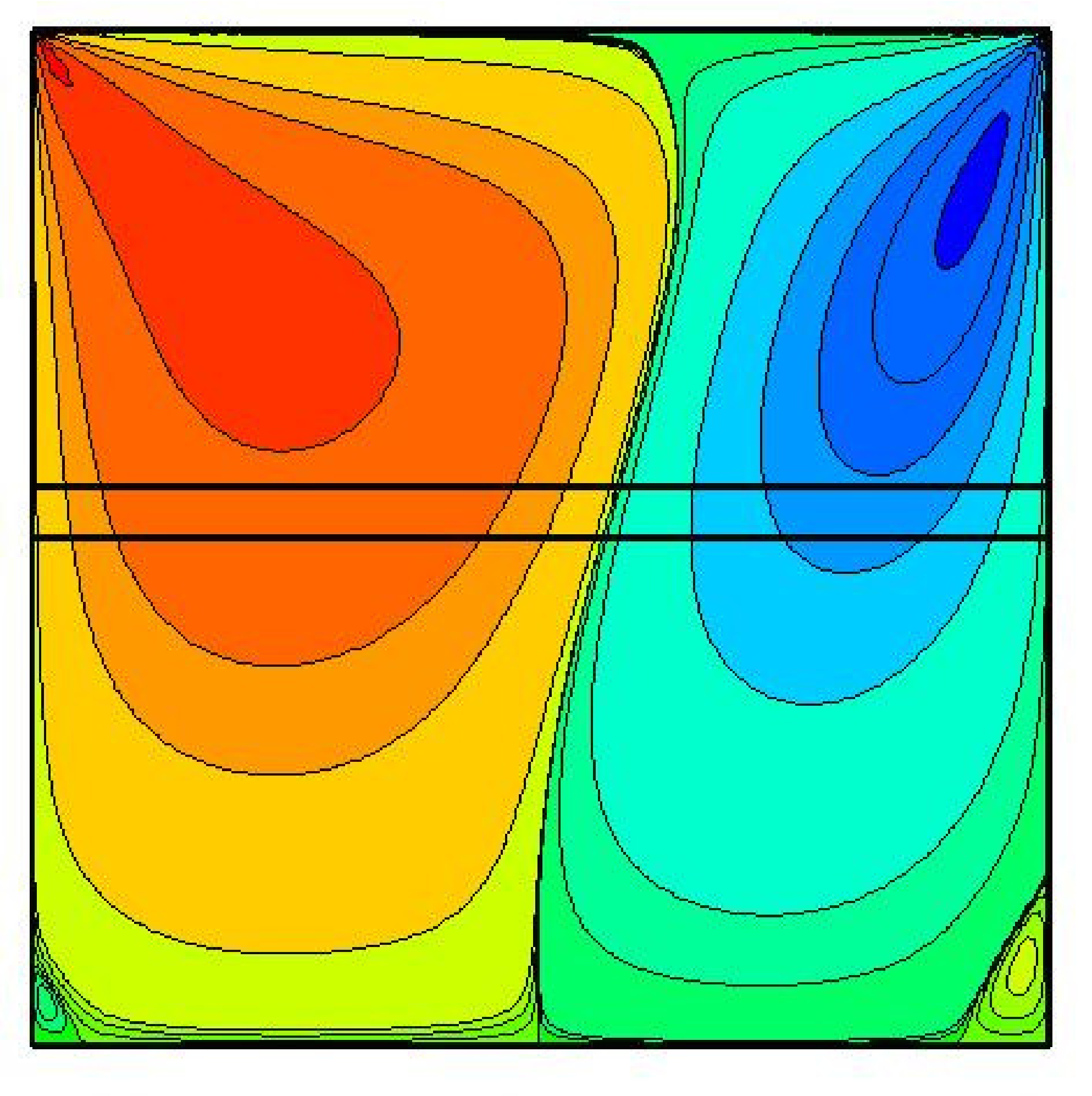}}
\scalebox{0.2}[0.2]{\includegraphics{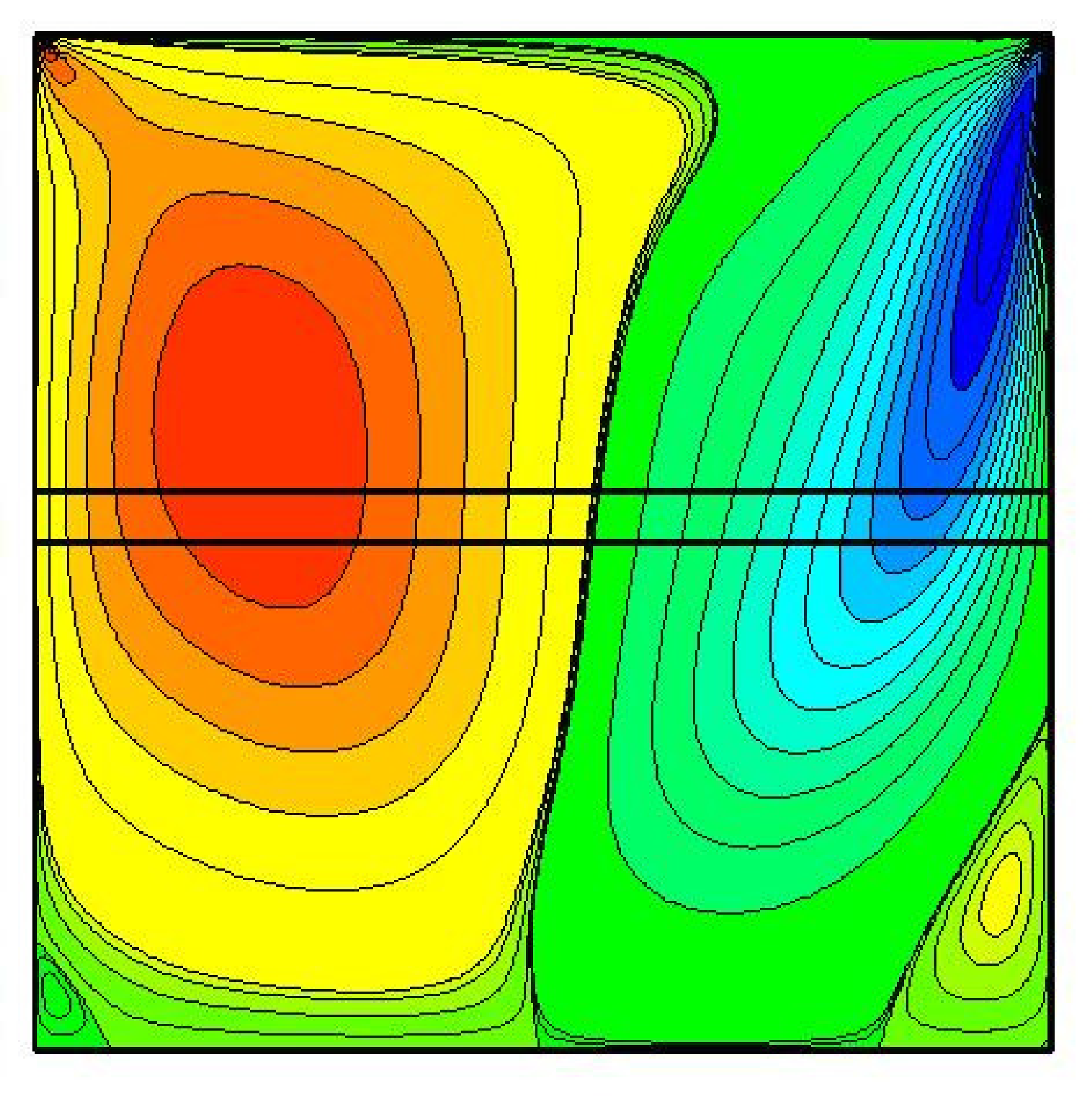}}
\scalebox{0.2}[0.2]{\includegraphics{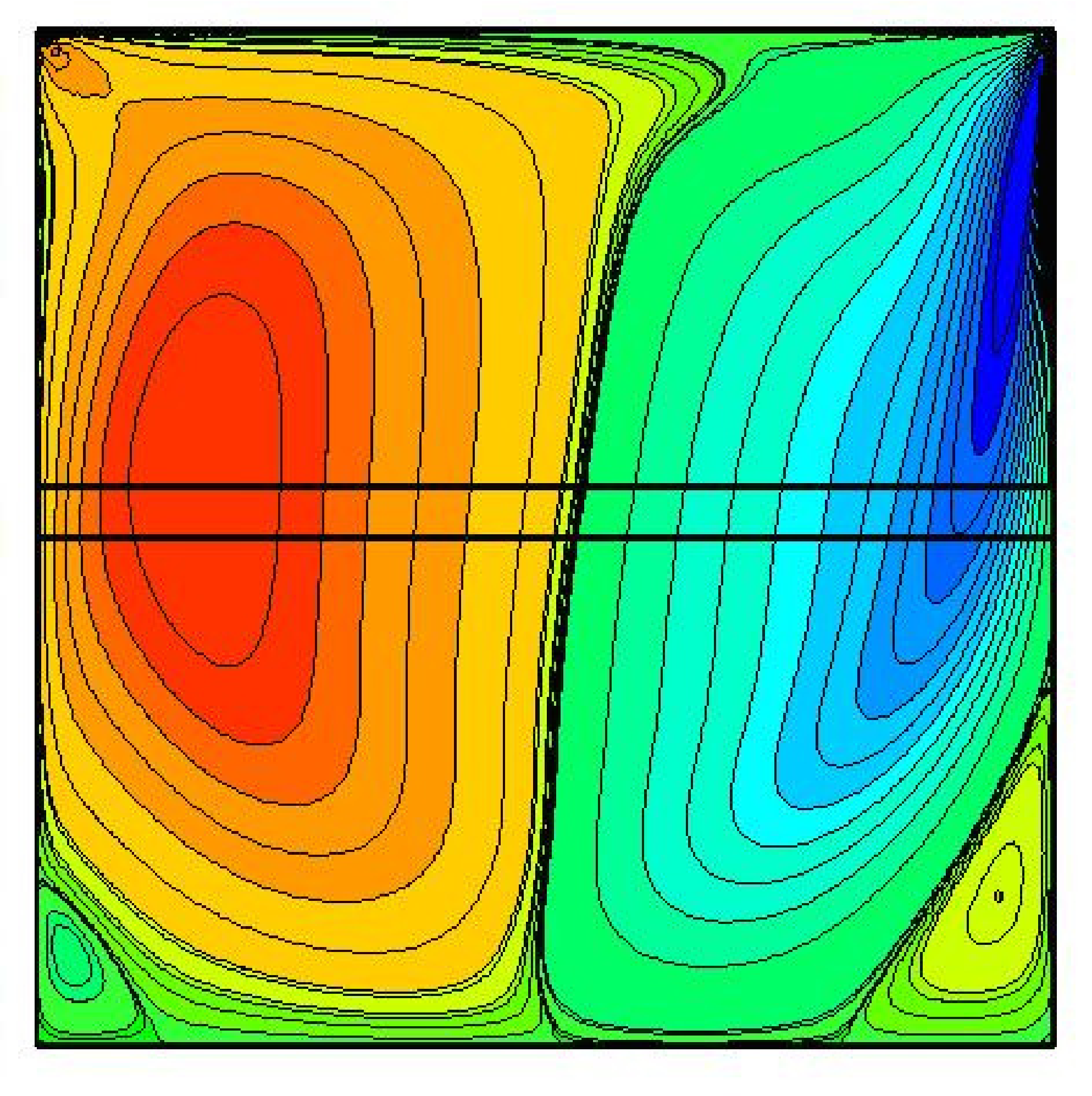}}\center{(a)
$Re=100$ \qquad\qquad(b) $Re=400$ \qquad\qquad(c) $Re=1000$}
\caption{ Contour plots of vertical velocity for different Reynolds
numbers} \label{fig.12}
\end{figure}

\begin{figure}
\centering
\scalebox{0.2}[0.2]{\includegraphics{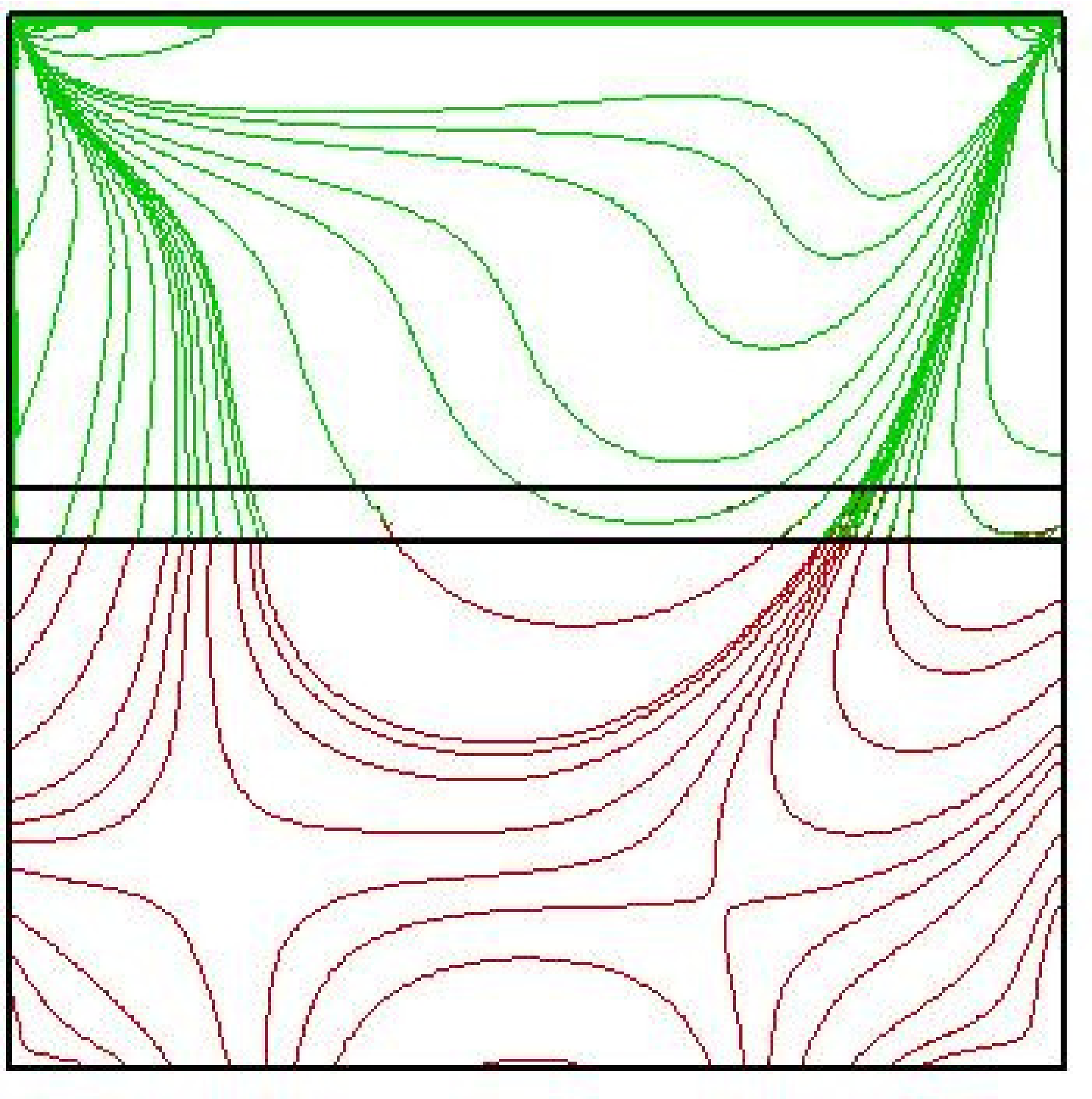}}
\scalebox{0.2}[0.2]{\includegraphics{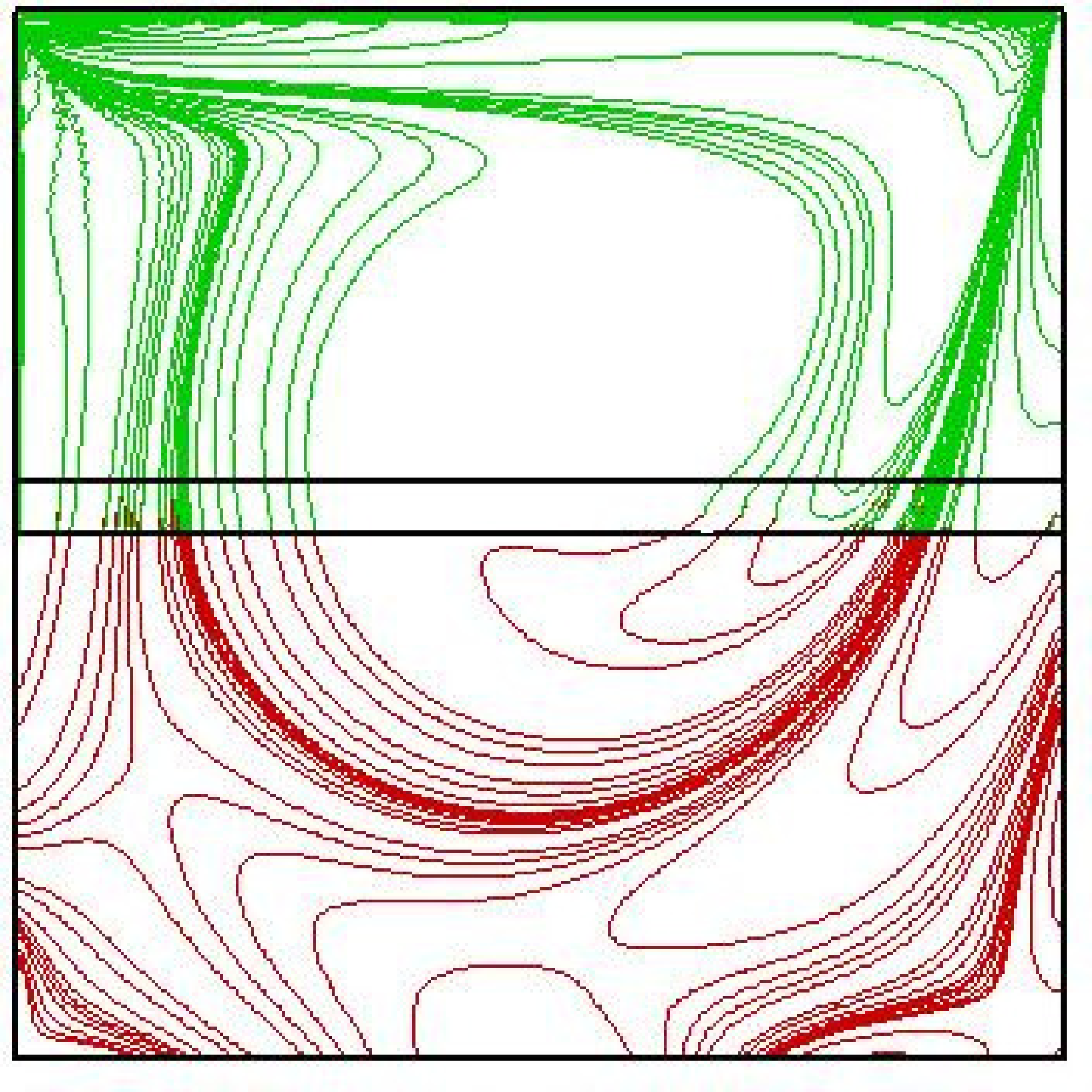}}
\scalebox{0.2}[0.2]{\includegraphics{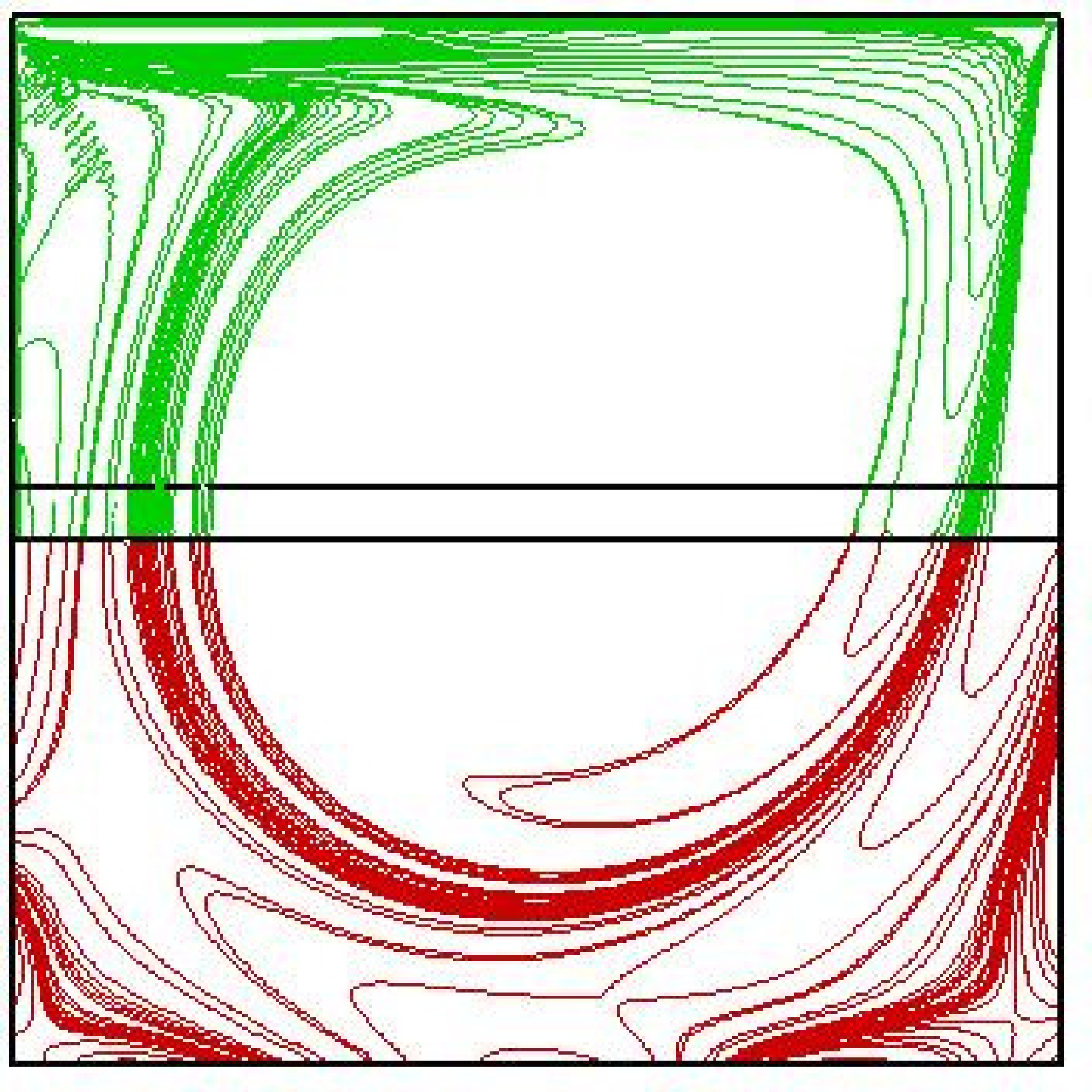}}\center{(a)
$Re=100$ \qquad\qquad(b) $Re=400$ \qquad\qquad(c) $Re=1000$}
\caption{ Contour plots of vorticity for different Reynolds numbers}
\label{fig.13}
\end{figure}
\end{document}